\documentclass[11pt,a4paper,english]{article}
\usepackage{jheppub}
\usepackage[T1]{fontenc}
\hypersetup{breaklinks=false,colorlinks=true,linkcolor=blue,citecolor=purple,pdftitle={Boosting Asymmetric charged DM via Thermalization}}
\usepackage{units,slashed,braket,physics,appendix,tikz}
\usepackage[compat=1.1.0]{tikz-feynman}
\tikzfeynmanset{warn luatex=false}

\title{Boosting Asymmetric Charged DM via Thermalization}
\author{Michael Geller}
\author{and Zamir Heller-Algazi}
\affiliation{School of Physics and Astronomy, Tel-Aviv University,\\Tel-Aviv 69978, Israel}
\emailAdd{mic.geller@gmail.com}
\emailAdd{zamir.heller@gmail.com}
\abstract{We consider a dark sector scenario with two dark matter species with opposite dark $U\pqty{1}$ charges and an asymmetric population comprising some fraction of the dark matter abundance. A new mechanism for boosting dark matter is introduced, arising from the large mass hierarchy between the two particles. In the galaxy, the two species thermalize efficiently through dark Rutherford scattering greatly boosting the lighter dark matter particle, far above the virial and escape velocities in the galaxy, while the dark charge prevents it from escaping. We study the consequences of this scenario for direct-detection experiments, assuming a kinetic mixing between the dark photon and the photon. If the charged dark sector makes up 5\% of the total DM mass in our galaxy and the mass ratio is between $10^3$--$10^4$, we find that current and future experiments may probe the boosted light dark matter for masses down to \unit[100]{keV}, in a hitherto unexplored parameter range.}
\arxivnumber{2210.03126}
\begin{document}
\maketitle
\flushbottom
	\section{Introduction}
	
	Astrophysical and cosmological observations of galaxy rotation curves~\cite{Persic:1995ru}, colliding galaxy clusters~\cite{Clowe:2003tk} and the Cosmic Microwave Background (CMB)~\cite{Planck:2018vyg} all indicate that a significant portion of matter in the Universe is not made of baryonic matter, the particles of the Standard Model (SM). Thus far, the missing \textit{Dark Matter} (DM) has only been observed indirectly by its gravitational influence on baryons, indicating that its non-gravitational interactions with the SM are, at most, very weak. Its underlying nature remains mostly unknown; we do not know its mass, whether it's made up of a single or of several different DM species, or how it interacts with itself or with baryonic matter.
	
	Results from various measurements and simulations point towards several noteworthy astrophysical properties of DM: it forms a halo around our galaxy whose profile is usually parameterized as NFW~\cite{Nesti:2013uwa,Navarro:1995iw} (other parameterizations are Burkert~\cite{Nesti:2013uwa,Burkert:1995yz,Salucci:2007tm} and Einasto~\cite{Sesar:2010fv,Navarro:2008kc,1965TrAlm...5...87E}) with a local density of $\rho_{\textrm{DM}}\simeq\unit[0.3]{GeV/cm^{3}}$~\cite{Read:2014qva} at the Earth's galactic location. The DM velocity distribution is usually taken as a Maxwellian distribution, truncated by the galactic escape velocity $v_\textrm{esc}\sim10^{-3}c$~\cite{Drukier:1986tm}, which produces an approximately constant velocity rotation curve as seen in observations.
	
	In the last few decades the leading DM candidate has been the \textit{Weakly Interacting Massive Particle} (WIMP), a single weak-scale particle weakly-coupled to the SM. For ${\cal O}\pqty{\unit{TeV}}$ scale masses and couplings of the same order as the weak force, the correct relic abundance is reproduced in what has been colloquially referred to as the ``WIMP miracle''. Such particles are predicted by several well-motivated theories beyond the Standard Model (BSM), most notably in supersymmetry (see for example~\cite{Fox:2019bgz} for a review on WIMPs and the minimal supersymmetric Standard Model). On the experimental front --- state of the art DM \textit{direct-detection} experiments are taking data, with the main effort focused on searching for the nuclear recoil from a rare scattering of DM against some target material. The current nuclear-recoil direct-detection experiments are XENON1T~\cite{XENON:2019gfn,XENON:2018voc}, PICO~\cite{PICO:2019vsc}, CRESST~\cite{CRESST:2019jnq}, DarkSide-50~\cite{DarkSide:2018bpj}, CDMSlite~\cite{SuperCDMS:2018gro}, PandaX-4T~\cite{Liu:2022zgu} and LZ~\cite{LZ:2022ufs}, and more are planned for the near future, most notably XENONnT~\cite{XENON:2020kmp}, SuperCDMS~\cite{SuperCDMS:2016wui} and DAMIC-M~\cite{Settimo:2020cbq}. This detection method is sensitive to heavier DM, in the WIMP range, while other experiments and analyses are based on electron recoil sensitive to lighter DM masses: XENON10~\cite {Essig:2017kqs}, SENSEI~\cite{SENSEI:2020dpa}, DAMIC~\cite{DAMIC:2019dcn}, DarkSide-50~\cite{DarkSide:2018ppu}, CDMSHVeV~\cite{SuperCDMS:2018mne} and EDELWEISS~\cite{EDELWEISS:2020fxc}. For even lighter masses, a detection strategy utilizing the wave nature of DM~\cite{Hook:2018dlk} is employed in CAST~\cite{CAST:2017uph}, ADMX~\cite{ADMX:2009iij,ADMX:2019uok}, CASPEr~\cite{Wu:2019exd}, MADMAX ~\cite{MADMAX:2019pub}, IAXO~\cite{IAXO:2019mpb} and ABRACADABRA~\cite{Salemi:2019xgl}.
	
	The immense experimental effort of searching for WIMPs has resulted in no discovery, and a significant fraction of the WIMP parameter space has already been excluded. At this stage, it is therefore important to consider broader ideas for DM beyond the standard paradigms. One such idea is the \textit{boosted DM} scenario proposed in~\cite{Agashe:2014yua}, wherein the DM is boosted to velocities significantly higher than its virial velocity $\pqty{\sim10^{-3}c}$. In \cite{Agashe:2014yua} a second subdominant and lighter component of DM is introduced. The massive species annihilate to the lighter species, transferring their rest-mass energy into kinetic energy, and as a result the lighter species are produced at high velocities. The higher momentum of the lighter species allows them to pass the detection threshold in direct-detection, enabling these experiments to probe mass ranges typically outside their reach. Other mechanisms of boosting DM were studied in~\cite{Bringmann:2018cvk,Ema:2018bih,Cappiello:2019qsw,An:2017ojc,Emken:2017hnp,Yin:2018yjn,Herrera:2021puj}.
	
	In this work we consider a new type of boosted dark matter, based on thermalization instead of the annihilation in~\cite{Agashe:2014yua}. Here, the dark sector will have a dark $U\pqty{1}_D$ gauge group and two fermions of opposite charges, a heavy $X$ and a light $\ell$. We assume an asymmetry in the dark sector, so that the fermion anti-particles are not present while the total charge is still zero because the fermion particles have opposite charge (similarly to the electron and the proton in the SM). The two components couple through dark-electric interactions and can therefore thermalize so that by the equipartition theorem, the light $\ell$ will be boosted to have similar kinetic energy as the heavy $X$. If the mass hierarchy of the light and heavy particles is large enough, the light DM may reach speeds far above the escape velocity while the dark-electric interaction with the heavy component will keep it bound to our galaxy. The cosmological history and possible signatures of a similar setting was analyzed in~\cite{Chacko:2018vss,Chacko:2021vin} within the context of the Mirror Twin Higgs model~\cite{Chacko:2005pe}. 
 
    Our dark sector interacts with the SM through a small kinetic mixing of the dark and SM photons. The kinetic mixing naturally induces an electric \textit{milli}-charge on the DM~\cite{Holdom:1985ag} while its dark EM charges are $\sim\mathcal{O}\pqty{1}$ (see~\cite{Agrawal:2016quu,Liu:2019knx,Feng:2009mn,Ackerman:2008kmp,Fan:2013yva,Fan:2013tia,McCullough:2013jma,Agrawal:2017rvu} for examples of other studies of milli-charged DM). As in~\cite{Agashe:2014yua} the lighter but faster DM can pass the energy detection threshold of terrestrial experiments, and we will see that future SENSEI runs will effectively probe milli-charge values of boosted $\ell$ at a range of \unit[100]{keV}--\unit[10]{MeV} unconstrained by any current data.
	
	\section{The Setup}\label{Sec. Setup}

	For our boosted dark matter model we introduce a dark sector with a $U\pqty{1}_D$ gauge group kinetically mixed with the SM $U\pqty{1}_\textrm{EM}$. The dark photon $\gamma_D$ is assumed to be massless, and the dark sector is comprised of two Dirac fermions, denoted $X$ and $\ell$, with opposite dark charges $\pm 1$. They are singlets under all other gauge groups. The Lagrangian is
	\begin{equation}
	\begin{aligned}
	    \mathcal{L}&\supset-\frac{1}{4}F_{\mu\nu}F^{\mu\nu}
		-\frac{1}{4}F_{\mu\nu}^{\prime}F^{\prime\mu\nu}
		-\frac{\epsilon}{2}F^{\mu\nu}F_{\mu\nu}^{\prime}\\
		&+\bar{X}\pqty{i\slashed{\partial}-g_D\slashed{A}'-m_X}X
		+\bar\ell\pqty{i\slashed{\partial}+g_D\slashed{A}'-m_\ell}\ell,
	\end{aligned}
	\end{equation}
	where $A_\mu^{\pqty{\prime}}$ is the SM (dark) photon, $F_{\mu\nu}^{\pqty{\prime}}$ its field strength, and $m_i$ the mass of fermion $i$ $\pqty{i=X,\ell}$. $g_D$ is the dark elementary charge and $\epsilon$ is the mixing of the dark and SM photons. Assuming small $\epsilon$, we can change the basis of the Lagrangian up to $\mathcal{O}\pqty{\epsilon}$ to decouple the dark and SM photons, which induces an electric charge on the dark fermions:
	\begin{equation}\label{Eq. no mix Lagrangian}
	\begin{aligned}
	    \mathcal{L}&\supset-\frac{1}{4}F_{\mu\nu}F^{\mu\nu}-\frac{1}{4}F_{\mu\nu}^{\prime}F^{\prime\mu\nu}\\
	    &+\bar{X}\pqty{i\slashed{\partial}-g_D\slashed{A}'-eQ\slashed{A}-m_X}X
		+\bar\ell\pqty{i\slashed{\partial}+g_D\slashed{A}'+eQ\slashed{A}-m_\ell}\ell,
	\end{aligned}
	\end{equation}
	with $Q\equiv\sqrt{\epsilon^{2}\alpha_{D}/\alpha_{\textrm{{QED}}}}$ the
	effective EM charge of the DM and $\alpha_{D}=g_{D}^{2}/4\pi$ the
	$U\pqty{1}_D$ fine-structure constant. With a small $\epsilon$, therefore, $X$ and $\ell$ obtain milli-charges. We set $\alpha_{D}$ to $\alpha_{\textrm{QED}}$ from here on out for simplicity so $Q=\abs{\epsilon}$.
 
	We will further assume both milli-charged particles (MCPs) have an asymmetric abundance where $X$ and $\ell$ remain, while $\bar{X}$ and $\bar{\ell}$ were annihilated. This asymmetry is similar to the SM baryon asymmetry, and can be produced with any of the many proposed baryogenesis mechanisms (see e.g.~\cite{Cline:2018fuq} for a review). We will assume that the MCPs are thermalized via the dark-electric interactions, and that $X$ is heavy while $\ell$ is light and that their mass ratio is $m_X/m_\ell\gg1$. When this mass ratio is large enough, the temperature of the $X$ and $\ell$ particles in the halo will be larger than the binding energy of $\pqty{X\ell}$ bound states, and hence the dark bound states which were formed during ``dark'' recombination are re-ionized. The binding energy of hydrogen-like atoms is $B_{X\ell}=m_\ell \alpha_D^2/2$ while the temperature is $T_X=m_Xv_X^2/2$, where $v_X\sim10^{-3}$ is roughly the virial velocity in the Milky Way, so requiring that $T_X\gtrsim B_{X\ell}$ results in $m_X/m_\ell\gtrsim10^2$. In this work we will consider larger mass ratios $m_X/m_\ell=10^3,10^4$ such that the dark plasma can be assumed to be completely ionized.
    
    While there are strong constraints on the magnitude of DM self-interactions~\cite{Feng:2009mn,Ackerman:2008kmp,DAmico:2009tep,Randall:2008ppe}, the constraints disappear if the self-interacting component is less than 10\% of the DM mass~\cite{Fan:2013yva,Chacko:2018vss}. We will assume for the rest of the paper that the MCP plasma makes up 5\% of the DM by mass. Our model also implicitly contains a standard cold dark matter (CDM) candidate $\chi$, neutral under $U\pqty{1}_D$, that accounts for most of the DM in the galaxy. We remain agnostic about its exact nature and the early universe dynamics which produce the abundances of the CDM and the charged dark sector. It should also be noted that, while there are claims that MCPs are evacuated from the galactic disk by supernova shock waves and prevented from returning to it by magnetic fields~\cite{Chuzhoy:2008zy,McDermott:2010pa}, this is not the case for our model since any energy gained by shocks is quickly turned into heat via their self-interactions~\cite{Foot:2010yz,Dunsky:2018mqs}.
    
    The constraints and new signatures arising from the cosmological history in a similar scenario were analyzed in~\cite{Chacko:2018vss}, where the equivalent $X$ and $\ell$ are the mirror proton and the mirror electron of the Mirror Twin Higgs model~\cite{Chacko:2005pe}, and the $U\pqty{1}_D$ is mirror electromagnetism. A viable cosmological history can be accommodated, with the main constraints stemming from the dark $U\pqty{1}_D$ contribution to $N_\textrm{eff}$. Additionally, new signatures arise due to the dark acoustic oscillations, which arise similarly to regular baryon acoustic oscillations. The oscillatory evolution of the density perturbation modifies the matter power spectrum on large scales, with the main effect of suppressing structure formation. A comprehensive study of the cosmological signatures of this model was performed in~\cite{Bansal:2022qbi}. 
    
    Another interesting phenomena to consider is plasma instability, which can occur in much of the parameter space of MCPs and $U\pqty{1}_D$ mediators, including the massless dark photon with strong self-interactions in our model~\cite{Ackerman:2008kmp,Lasenby:2020rlf,Heikinheimo:2015kra}. However, although the plasma almost certainly exhibits instability, its observational effects are poorly understood due to the highly nonlinear nature of instabilities. Simulations of dark plasma instabilities have been performed previously~\cite{Spethmann:2016glr} in the context of galaxy cluster collisions, and may offer a viable explanation for some features of the DM distribution of some observations. Dark plasma instabilities offer an intriguing avenue of exploration for dark photon models, with potentially huge constraining power in much of the parameter space, but would require detailed numerical studies and simulations to make concrete statements about said models. We ignore these complexities here, and therefore we will refrain from deriving robust bounds on our scenario in this work. 
	
	We summarize the dark sector particle content in table~\ref{Table DM Assignments}.
	\begin{table}
		\centering
		\begin{tabular}{|c|c|c|c|} 
			\hline
			DM  & mass & $U\pqty{1}_D$ & $U\pqty{1}_\textrm{EM}$ \\ [0.5ex] 
			\hline
			$X$ & $10^{-1}$--$\unit[10]{GeV}$ & $+1$ & $+Q$ \\ [0.5ex] 
			
			$\ell$ & $10^{-1}$--$\unit[10]{MeV}$ & $-1$ & $-Q$ \\ [0.5ex] 
			
			$\chi$ & -- & 0 & 0 \\ [0.5ex] 
			\hline
		\end{tabular}
		\caption{Masses and charges of the dark sector particles $X$, $\ell$ (the MCPs) and $\chi$ (the unspecified CDM).}\label{Table DM Assignments}
	\end{table}
	
	\section{Galactic MCP Distribution}\label{Sec. Galactic MCP Distribution}
	
	In order to study the potential for detection of the boosted MCPs on Earth, we first analyze their distribution in our galaxy. The charged dark sector has a subdominant mass fraction with little effect on the dominant neutral CDM distribution, for which we take a standard NFW profile $\rho_{\textrm{NFW}}\pqty{r}$ in the Milky Way~\cite{Nesti:2013uwa}. For the analysis here we assume the $X$ and $\ell$ plasmas are thermalized and in Sec.~\ref{Sec. l-X Thermalization} we will find the parameter space wherein this assumption holds. The typical mean velocity of the $X$ particles is of order $v_X\sim10^{-3}$, the virial velocity induced by the dominant CDM potential. The hierarchical mass ratio between the massive $X$ and light $\ell$ results in a significant boost of the $\ell$ particles, possibly to speeds above the escape velocity. We found in an explicit calculation that the $\ell$ particles are still bound to the galaxy by their $U\pqty{1}_D$ interaction with the $X$ particles. We model the halo dynamics of the $\ell$ and $X$ plasmas as ideal gases in thermal and hydrostatic equilibria, with the neutral CDM acting as a background gravitational potential, and neglect dissipation processes in the MCP plasma. We further discuss dissipation, and the regions of parameter space where it is important, in Sec.~\ref{Sec. Cooling}. Our analysis assumes spherical symmetry, and we leave the study of more complicated magneto-hydrodynamic effects that may break this symmetry for a future analysis. 

	The pressure $P_i$ of each MCP plasma component balances the gravitational and dark-electric forces acting on it. The hydrostatic equilibrium condition of each component and Gauss's law for the dark charge and gravity is described by the following set of equations:
	\begin{equation}\label{Eq. All Hydrostatic Equations}
		\begin{aligned}
			\dv{P_\ell}{r} & =-n_{\ell}\alpha_{D}\frac{Q_\textrm{tot}\pqty{r}}{r^{2}}
			-m_\ell n_\ell \frac{GM_\textrm{tot}\pqty{r}}{r^2}, &
			Q_\textrm{tot}\pqty{r}& =\int_{0}^{r}\pqty{n_{X}-n_{\ell}}4\pi r^{2}\dd{r},\\
			\dv{P_{X}}{r}& =n_{X}\alpha_{D}\frac{Q_\textrm{tot}\pqty{r}}{r^{2}}-m_{X}n_{X}\frac{GM_\textrm{tot}\pqty{r}}{r^{2}},
			& M_\textrm{tot}\pqty{r}& =\int_{0}^{r}\pqty{m_{X}n_{X}+m_\ell n_\ell+\rho_{\textrm{NFW}}}4\pi r^{2}\dd{r},
		\end{aligned}
	\end{equation}
	where $Q_\textrm{tot}$ $\pqty{M_\textrm{tot}}$ is the total dark charge (mass) of the halo. The SM electric force contribution is negligible since the MCP's SM charge is suppressed by $\epsilon$ with respect to the dark charge. The CDM component dominates the halo mass $M_\textrm{tot}\pqty{r}\simeq M_{\textrm{CDM}}\pqty{r}$ and hence we neglect the MCP's contribution to the gravitational potential $\Phi$, given therefore by $\nabla\Phi=GM_\textrm{CDM}/r^2$.
	
	Due to the long-range dark electric force, the $X$ and $\ell$s are separated from each other only by a Debye length, which is much smaller than the length scale of the galaxy. Thus the number densities of the light and heavy particles trace each other, effectively setting $n_{X}\simeq n_{\ell}\pqty{\equiv n_{\textrm{MCP}}}$. This is the neutral plasma approximation, and we can use it to simplify Eq.~\eqref{Eq. All Hydrostatic Equations} to a single hydrostatic equilibrium condition,
	\begin{equation}\label{Eq. Single Hydrostatic Equation}
		\dv{\pqty{P_X+P_\ell}}{r}=-\rho_{\textrm{MCP}}\nabla\Phi,
	\end{equation}
	where $\rho_{\textrm{MCP}}=m_Xn_X+m_\ell n_\ell\simeq m_Xn_{\textrm{MCP}}$ is the total MCP mass density. We have self-consistently checked that $Q_\textrm{tot}\ll1$ in our solution by extracting $Q_\textrm{tot}\pqty{r}$ from Eq.~\eqref{Eq. All Hydrostatic Equations}.
	
	We now solve Eq.~\eqref{Eq. Single Hydrostatic Equation} to find $\rho_{\textrm{MCP}}$, using similar methods to those used in \cite{Chacko:2021vin}. To fully solve the equation, we need to specify the temperature profile of the halo. As we show in Sec.~\ref{Sec. Thermal Conduction} thermal conduction is effective in the relevant range of parameters, and the MCP halo is approximately isothermal. Using plasma neutrality and the ideal gas equation-of-state $P_i=n_iT_i$ we find that the total pressure of the MCP plasma is twice the pressure of the $X$ species, $P_X+P_\ell=2P_X\simeq2\rho_{\textrm{MCP}}\theta_X$ (where $\theta_i\equiv T_i/m_i$ is the temperature divided by the mass). Then the density profile is given by
	\begin{equation}
		2\theta_{X}^\textrm{iso}\dv{\rho_{\textrm{MCP}}^{\textrm{iso}}}{r}
		=-\rho_{\textrm{MCP}}^{\textrm{iso}}\dv{\Phi}{r}
		\implies\rho_{\textrm{MCP}}^{\textrm{iso}}\pqty{r}
		=\rho_{\textrm{MCP}}^{\textrm{iso}}\pqty{0}
		\exp\bqty{\frac{\Phi\pqty{0}-\Phi\pqty{r}}{2\theta_{X}^\textrm{iso}}},
	\end{equation}
	where $\Phi$ is the gravitational potential of the CDM component in a NFW profile. The central density $\rho_{\textrm{MCP}}^{\textrm{iso}}\pqty{0}$ is set by requiring that the total fractional MCP mass in the galaxy is $f_\textrm{MCP}\equiv M_{\textrm{MCP}}/M_{\textrm{CDM}}$, which we will take as 5\% in this paper. It is trivial to rescale the results of this paper for any other value of $f_\textrm{MCP}$.
	
	To find the temperature $\theta_X^\textrm{iso}$, we assume that the initial conditions, and in particular, the total initial energy of the MCP halo, are independent of the $U\pqty{1}_D$ interaction. Therefore, to find this initial energy, we consider the hypothetical scenario where these interactions are turned off. Then the ``collisionless'' MCPs would be indistinguishable from the CDM component and would follow the same profile shape. Taking the usual assumption that the system is virialized we may use the virial theorem $2K_{\textrm{vir}}+U_\textrm{vir}=0$ to relate the average kinetic energy of the particles to their potential energy. For this, we define the virial temperature as
	\begin{equation}\label{Eq. Virial CDM Temperature}
		\theta_{\textrm{CDM}}\equiv\frac{2}{3}\frac{K_{\textrm{CDM}}}{M_{\textrm{CDM}}}
		=\frac{1}{3}\frac{\left|U_{\textrm{CDM}}\right|}{M_{\textrm{CDM}}}
		=\frac{\int GM_{\textrm{CDM}}\rho_{\textrm{NFW}}4\pi r \dd{r}}{3M_{\textrm{CDM}}}.
	\end{equation}
	The total energy is then
	\begin{equation}\label{Eq. Virial CDM Energy}
		E_{\textrm{vir}}=U_{\textrm{vir}}+K_{\textrm{vir}}=-K_{\textrm{vir}}
		=-\frac{3}{2}M_{\textrm{MCP}}\theta_{\textrm{CDM}}.
	\end{equation}
	In our scenario, the collisions of the MCPs redistribute this initial energy among themselves but the total energy is conserved. The potential energy of the MCP halo is
	\begin{equation}\label{Eq. Isothermal Potential Energy}
		U_{\textrm{iso}}=\int\Phi\rho_{\textrm{MCP}}^{\textrm{iso}}4\pi r^{2}\dd{r}.
	\end{equation}
	The total kinetic energy of the isothermal MCP halo is\footnote{The coefficient $3/2$ is inaccurate for relativistic gases like the $\ell$ plasma, and approaches $3$ in the ultra-relativistic limit. Nevertheless, in our parameter space of interest the relativistic correction to this coefficient is negligible, see Eq.~\eqref{Eq. Average Relativisitc Kinetic Energy}.}
	\begin{equation}\label{Eq. Isothermal Kinetic Energy}
		K_{\textrm{iso}}=\int\frac{3}{2}\left(n_{X}+n_{\ell}\right)T\dd{V}
		=\int3\rho_{\textrm{MCP}}^{\textrm{iso}}\theta_{X}^\textrm{iso}\dd{V}
		=3\theta_{X}^\textrm{iso}M_{\textrm{MCP}}.
	\end{equation}
	We find $\theta_X^\textrm{iso}$ by requiring energy conservation
	\begin{equation}\label{Eq. Isothermal Halo Energy Conservation}
		E_{\textrm{iso}}=K_{\textrm{iso}}+U_{\textrm{iso}}=E_{\textrm{vir}}.
	\end{equation}
	
\begin{figure}
	   \begin{centering}
		\includegraphics[width=0.6\columnwidth]{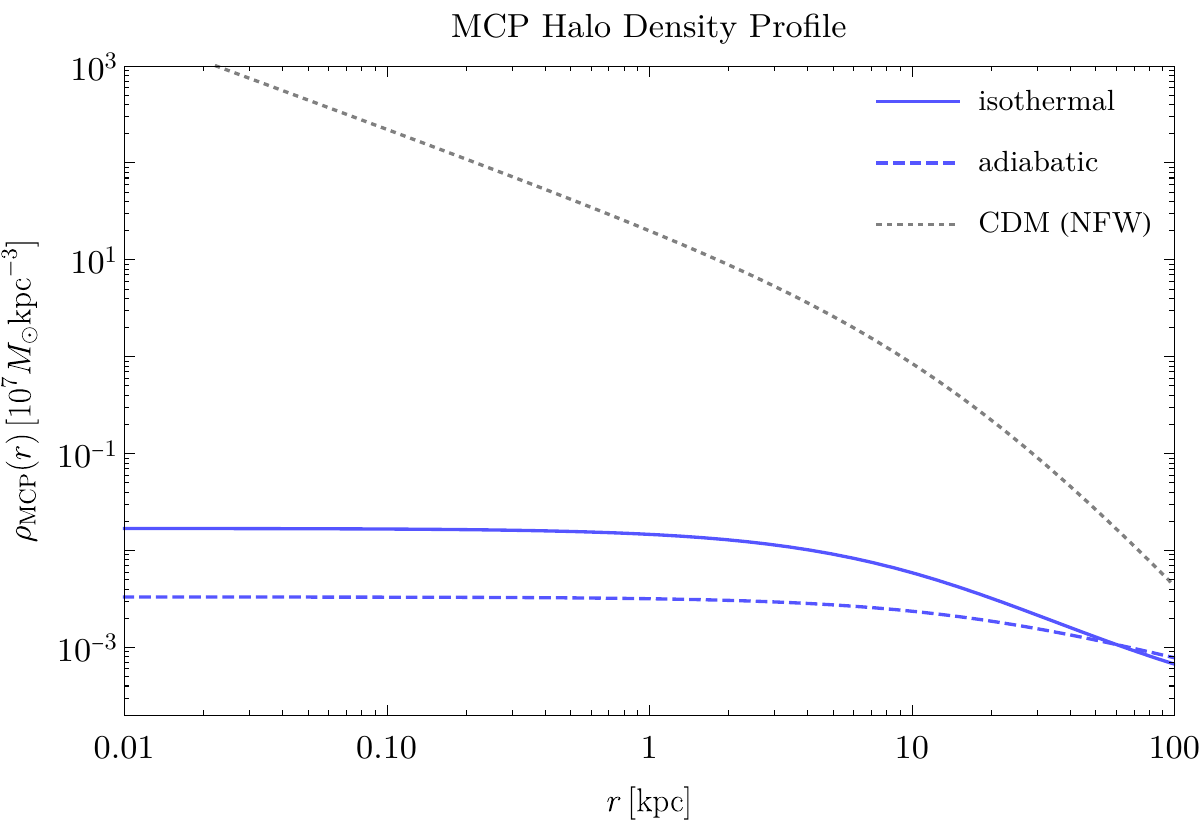}
		\par\end{centering}
	   \caption{The mass density $\rho_{\textrm{MCP}}\pqty{r}$ inferred from hydrostatic equilibrium for an isothermal and adiabatic halo, with $\rho_{\textrm{CDM}}\pqty{r}$ for comparison. The total fractional MCP mass was set to be $f_\textrm{MCP}=5\%$. The exact forms of both solutions are derived in Appendix~\ref{App. MCP Halo Analytic Results}. \label{Fig. MCP Mass-density Profiles}}
    \end{figure}

	This fixes the MCP's mass density profile, which is shown in Fig.~\ref{Fig. MCP Mass-density Profiles}. In similar fashion one can also solve the case where heat conduction is inefficient and heat convection results in an adiabatic halo, whose density profile is also shown in Fig.~\ref{Fig. MCP Mass-density Profiles}. In Appendix~\ref{App. MCP Halo Analytic Results} we present the exact analytical solutions calculated in this section as well as explicitly solve the hydrostatic equilibrium condition in Eq.~\eqref{Eq. Single Hydrostatic Equation} for the adiabatic halo. In the following, unless stated otherwise, the MCP number density and temperature are computed in the isothermal halo approximation.

	\section{Thermodynamic Considerations}\label{Sec. Thermodynamic Considerations}
	
	In the previous section we calculated the MCP profile assuming thermal and hydrostatic equilibria. Here we investigate some of the assumptions we have previously made. Firstly, we have assumed the $\ell$ and $X$ plasmas are thermalized, $T_\ell=T_X$. This is determined by the thermalization rate between the two species. Secondly, we have taken the MCP halo to be isothermal. This assumption depends on the rate of heat transfer in the halo. Lastly we have neglected dissipation processes through which the MCP plasma is cooled that would drive the collapse of the MCP halo to discs and other possible structures. Such cooling effects can be ignored if their energy loss is small. Below, we examine the validity of all these assumptions, and find the parameter space in which they are justified. This parameter space is presented on the plot in Fig.~\ref{Fig. m_ell-Q Parameter Space}. We further expand on the analysis in this section in Appendix~\ref{App. Thermodynamic Regime Plots}.
	
	\subsection{$\ell-X$ Thermalization}\label{Sec. l-X Thermalization}
	
	We first analyze the rate of thermalization between the $\ell$ and $X$ components. The two plasmas exchange energy through repeated collisions of $\ell$ and $X$ particles until their average kinetic energies equalize. We compare the thermalization timescale $t_\textrm{th}$ through these collisions to the lifetime of our galaxy $t_\textrm{gal}$ in the parameter range of interest. We follow the results of~\cite{SS-thesis,1983MNRAS.202..467S,1985ApJ...295...28D,1986ApJ...307...47D} to compute the thermalization time, identifying the electrons with our light MCP $\pqty{\ell}$ and the protons with their heavy MCP partners $\pqty{X}$.
	
	The energy exchange rate per unit volume of $\ell$ particles due to Rutherford scattering with a non-relativistic $X$ is given by
	\begin{equation}
        P_\textrm{Rutherford}
		=-\frac{3m_{\ell}}{2m_{X}}\ln\Lambda
		\frac{e^{-1/\theta_{\ell}}\theta_{\ell}^{1/2}}{K_{2}\pqty{1/\theta_{\ell}}}
		\frac{1+2\pqty{\theta_{\ell}+\theta_{X}}+2\pqty{\theta_{\ell}+\theta_{X}}^2}{\pqty{\theta_{\ell}+\theta_{X}}^{3/2}}
		 n_Xn_\ell\sigma_T\pqty{T_{\ell}-T_{X}},
	\end{equation}
	where $\sigma_{T}\equiv8\pi\alpha_{D}^{2}/3m_{\ell}^{2}$ is the dark Thomson cross-section. $\ln\Lambda$ is the Coulomb logarithm, where $\Lambda\sim T_{\ell}/\omega_{p}$ with $\omega_{p}$ the relativistic plasma frequency $\omega_{p}^{2}=\frac{4\pi}{3}n_{\ell}\alpha_{D}/T_{\ell}$, the effective cutoff for the IR divergence of Rutherford scattering. 
	
	The relaxation frequency, the inverse of the thermalization time $t_\textrm{th}$, is given by
	\begin{equation}\label{Eq. Relaxation Frequency}
		\nu^{\ell X}=\abs{\frac{P_\textrm{Rutherford}}{\varepsilon n_\ell \pqty{T_\ell-T_X}}},
	\end{equation}
	where $\varepsilon$ is the average kinetic energy of $\ell$ particles normalized to their temperature,
	\begin{equation}\label{Eq. Average Relativisitc Kinetic Energy}
		\varepsilon\equiv\frac{\Braket{\gamma_\ell}-1}{\theta_\ell}=\frac{1}{\theta_{\ell}}\bqty{\frac{K_{1}\left(1/\theta_{\ell}\right)}{K_{2}\left(1/\theta_{\ell}\right)}+3\theta_{\ell}-1}
		\longrightarrow\begin{cases}
			\frac{3}{2}& \theta_\ell\ll1\\
			3 & \theta_\ell\gg1
		\end{cases}.
	\end{equation}
	The relaxation frequency is then
	\begin{equation}
		\nu^{\ell X}
		=\frac{1}{\varepsilon}\frac{3m_{\ell}}{2m_{X}}\ln\Lambda
		\frac{e^{-1/\theta_{\ell}}\theta_{\ell}^{1/2}}{K_{2}\pqty{1/\theta_{\ell}}}
		\frac{1+2\pqty{\theta_{\ell}+\theta_{X}}+2\pqty{\theta_{\ell}+\theta_{X}}^2}{\pqty{\theta_{\ell}+\theta_{X}}^{3/2}}
		n_X\sigma_T.
	\end{equation}

\begin{figure}
		\begin{centering}
			\includegraphics[width=0.5\columnwidth]{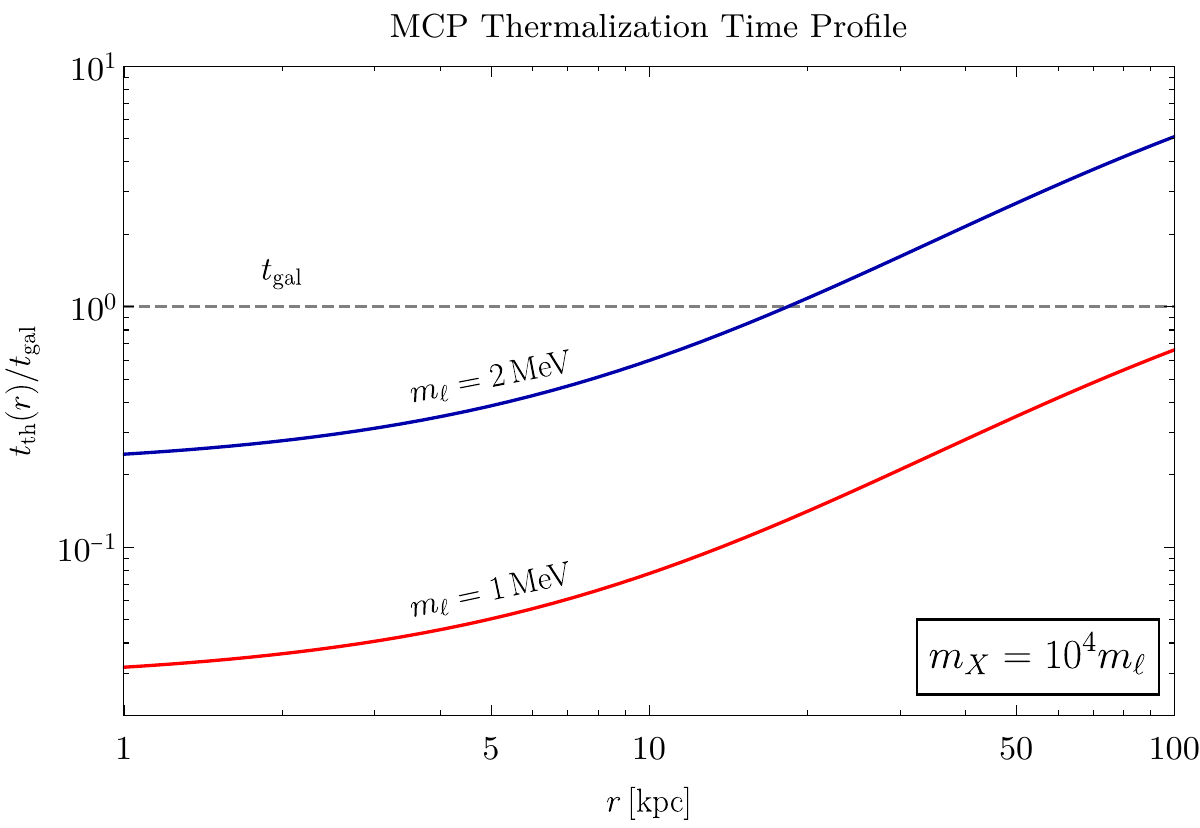}\includegraphics[width=0.5\columnwidth]{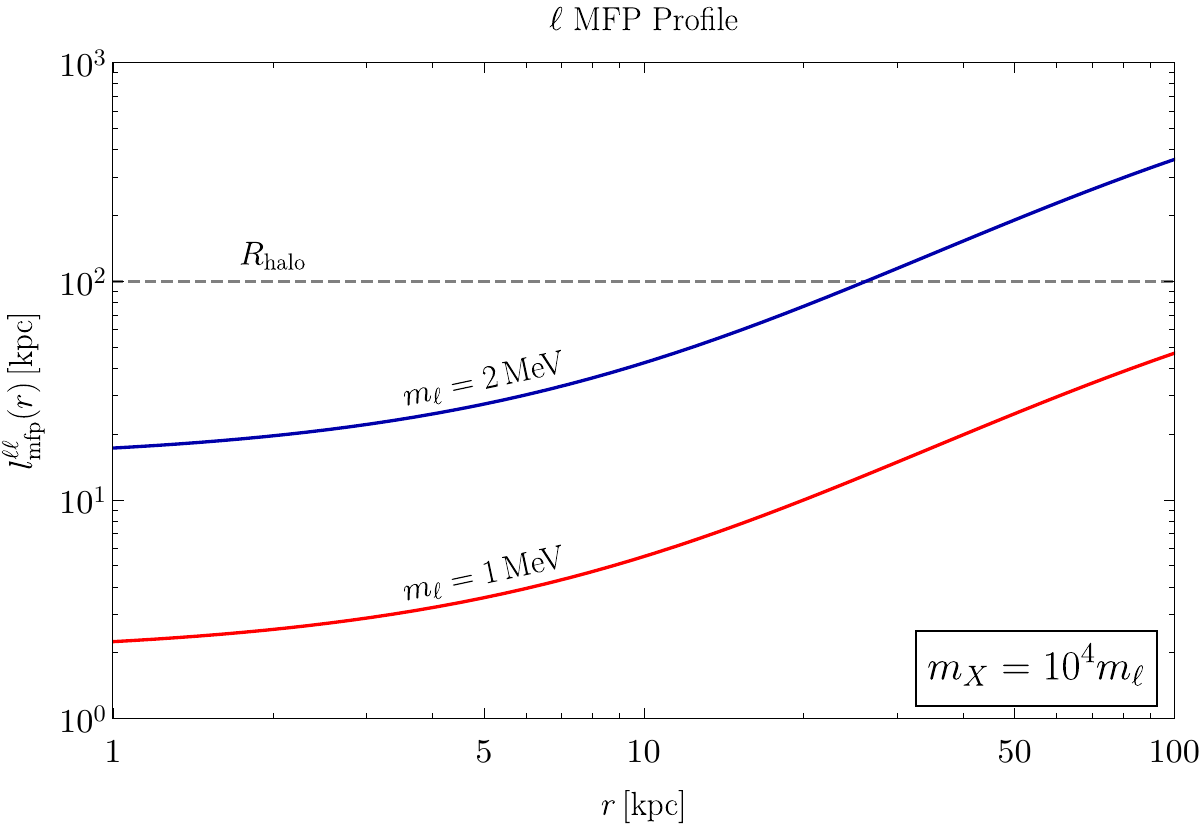}
			\par\end{centering}
		\caption{The thermalization timescale $t_\textrm{th}$ (left) and the $\ell$ mfp $l^{\ell\ell}_\textrm{mfp}$ (right) as a function of the halo radius $r$. $t_\textrm{th}$ is normalized by the galaxy age $t_\textrm{gal}$, and $l^{\ell\ell}_\textrm{mfp}$ is compared to the halo length $R_\textrm{halo}$.\label{Fig. Thermalization and l-l mfp Profiles}}
	\end{figure}

	We first note that the thermalization rate is dependent on $n_X$ and hence varies across the halo, with the denser inner regions thermalizing faster. The Solar System's location in the galaxy is in this inner region, and it is therefore possible that the two species are thermalized locally in our vicinity even if in the outer regions their temperatures differ, e.g. as in the blue curve in Fig.~\ref{Fig. Thermalization and l-l mfp Profiles} on the left. However, the mean-free-path (mfp) of the $\ell$ particles in the halo can be fairly large in this part of our parameter space (see Eq.~\eqref{Eq. l-l mfp} and Fig.~\ref{Fig. Thermalization and l-l mfp Profiles} on the right), making it harder to analyze the location from whence the $\ell$ particles arrive to the Earth. To simplify the discussion, we make the conservative assumption that the thermalization of $\ell$ and $X$ particles is determined by the rate averaged over the entire galaxy --- which is always slower than the local rate.

	To find the parameter space where our thermalization assumption is valid, we compare the thermalization timescale $t_\textrm{th}$ to the age of our galaxy $t_\textrm{gal}$. We find that for any choice of the ratio $m_{X}/m_{\ell}$  our assumption is valid for low enough values of $m_\ell$ and breaks down above some critical mass. Above this mass the two plasmas have not reached thermal equilibrium within the lifetime of the galaxy and have separate temperatures. In this scenario the $\ell$ particles can still be somewhat boosted by the heat transfer between the plasmas, but we leave the analysis of this case for a future work. In Fig.~\ref{Fig. m_ell-Q Parameter Space} we show these maximal allowed masses as solid purple lines for two choices of the mass ratio $m_X/m_\ell=\Bqty{10^3,10^4}$.
	
	\subsection{Thermal Conduction}\label{Sec. Thermal Conduction}
	
	To determine whether our assumption of an isothermal MCP halo is valid, we estimate the rate of thermal conduction in the galaxy. Heat diffuses from one region of space to another in a series of rapid particle collisions, exchanging energy in the process until they reach kinetic equilibrium. The heat flux density $\vb{q}$ is proportional to the temperature gradient $\vb{q}=-\kappa\grad{T}$ where $\kappa$ is the thermal conductivity coefficient. The effective thermal conductivity coefficient in plasma is given by~\cite{spitzerplasmabook}
	\begin{equation}
		\kappa_\textrm{eff}\simeq 0.225 \times 20 \pqty{\frac{2}{\pi}}^{3/2}\frac{8\pi}{3}\frac{\theta_\ell^{5/2}}{\sigma_T\ln\Lambda}.
	\end{equation}
	The internal heat energy density of the MCP plasma is $Q=\frac{5}{2}n_XT+\frac{5}{2}n_\ell T=5n_\textrm{MCP}T$.\footnote{Generally the internal heat density is $Q=\hat{c}_PnT$ where $\hat{c}_P$ is the specific heat capacity at constant pressure. For ideal monoatomic gases $\hat{c}_P=5/2$, where we neglect relativistic deviations to this coefficient for the $\ell$ plasma.} Energy conservation then implies the continuity equation $\partial_tQ+\div{\vb{q}}=0$, from which the  heat equation follows:
	\begin{equation}\label{Eq. Heat Equation}
		5\pqty{\pdv{n_\textrm{MCP}}{t}T+n_\textrm{MCP}\pdv{T}{t}}=\div(\kappa_\textrm{eff}\grad{T}).
	\end{equation}
	
    As the temperature and number density change in time so does the total pressure, $P_X+P_\ell=2n_\textrm{MCP}T=\frac{2}{5}Q$. The pressure fluctuations propagate at the speed of sound in the halo $v_s\sim\sqrt{P/\rho}\sim\sqrt{\theta_X}\sim v_X$ (since $X$ dominate the MCP density), and pass through the entire halo $\sim R_\textrm{halo}$ in about $\unit[0.5]{Gyr}\ll t_\textrm{gal}$.  Therefore, we demand hydrostatic equilibrium at any given time with a quasi-static evolution due to the thermal conduction. We can then differentiate Eq.~\eqref{Eq. Single Hydrostatic Equation} with respect to time,
	\begin{equation}\label{Eq. Hydrostatic Derivative}
		-m_X\pdv{n_\textrm{MCP}}{t}\grad{\Phi}=\grad{\pdv{\pqty{P_X+P_\ell}}{t}}=\frac{2}{5}\grad(\div(\kappa_\textrm{eff}\grad{T})).
	\end{equation}
	In the last equality we plugged in Eq.~\eqref{Eq. Heat Equation}. We solve Eqs.~\eqref{Eq. Heat Equation} and~\eqref{Eq. Hydrostatic Derivative} to obtain $\pdv*{T}{t}$, and find the thermal conduction timescale is given by
	\begin{equation}\label{Eq. Conduction Timescale}
		t_\textrm{cond}\equiv\abs{\frac{T}{\pdv*{T}{t}}}
		=\abs{\frac{5m_Xn_\textrm{MCP}T\grad{\Phi}}{m_X\grad{\Phi}\div(\kappa_\textrm{eff}\grad{T})+2\grad(\div(\kappa_\textrm{eff}\grad{T}))T}}.
	\end{equation}

\begin{figure}
		\begin{centering}
			\includegraphics[width=0.6\columnwidth]{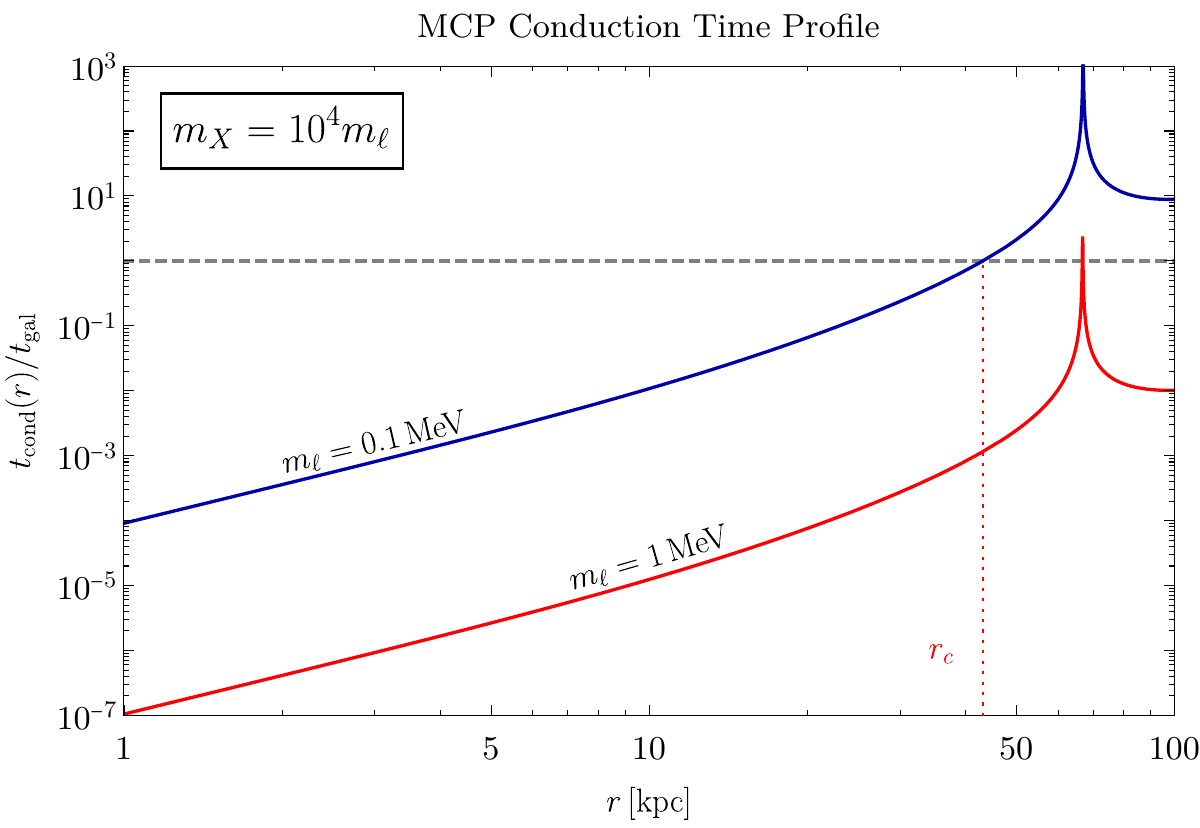}
			\par\end{centering}
		\caption{The conduction timescale $t_\textrm{cond}$ defined in Eq.~\eqref{Eq. Conduction Timescale} across the halo normalized by the galaxy age $t_\textrm{gal}$. The point at which $t_\textrm{cond}$ diverges is an artefact of an arbitrary choice in Eq.~\eqref{Eq. Conduction Timescale}: it is where $\pdv*{T}{t}$ vanishes for the \textit{initial} adiabatic halo. As the halo evolves the temperature profile changes with it, $\pdv*{T}{t}$ will no longer be zero and heat will begin to conduct there.\label{Fig. Conduction Profile}}
	\end{figure}

	As we now show, the MCP halo density and temperature are well approximated by an isothermal halo in our parameter range of interest, and the profile starts to deviate considerably from an isothermal distribution only when cooling also becomes important. To see this, we first note that in the absence of heat conduction, convection results in an adiabatic halo. To check whether thermalization occurs, we take this to be the halo's initial condition, and check if the rate of heat transfer according to Eq.~\eqref{Eq. Conduction Timescale} is fast enough to thermalize the halo within $t_\textrm{gal}$. This rate is position-dependent and, in particular, faster in the inner parts of the halo (see Fig.~\ref{Fig. Conduction Profile}). For low $m_\ell$ masses we find that heat transfer is inefficient throughout and the halo remains completely adiabatic, and for high $m_\ell$ masses heat transfer is efficient throughout and the entire halo is isothermal.

\begin{figure}
		\begin{centering}
			\includegraphics[width=0.488\columnwidth]{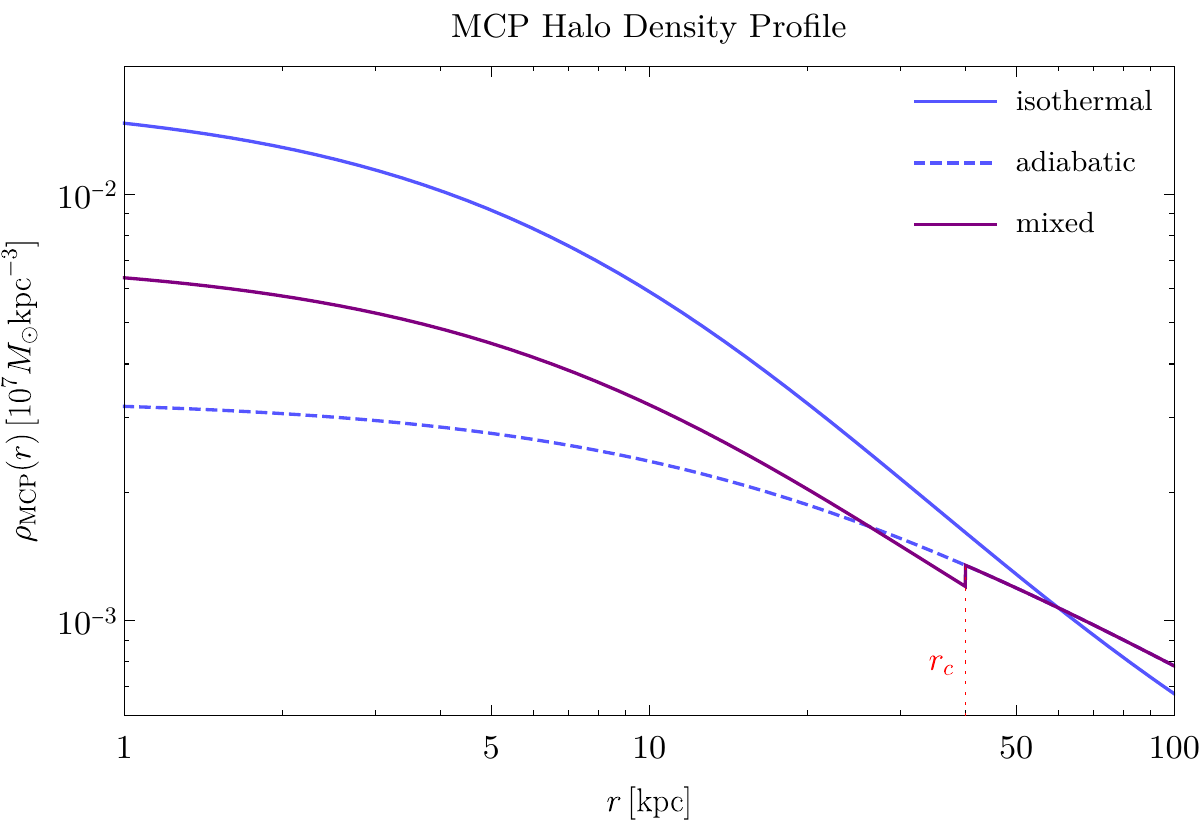}\includegraphics[width=0.5116\columnwidth]{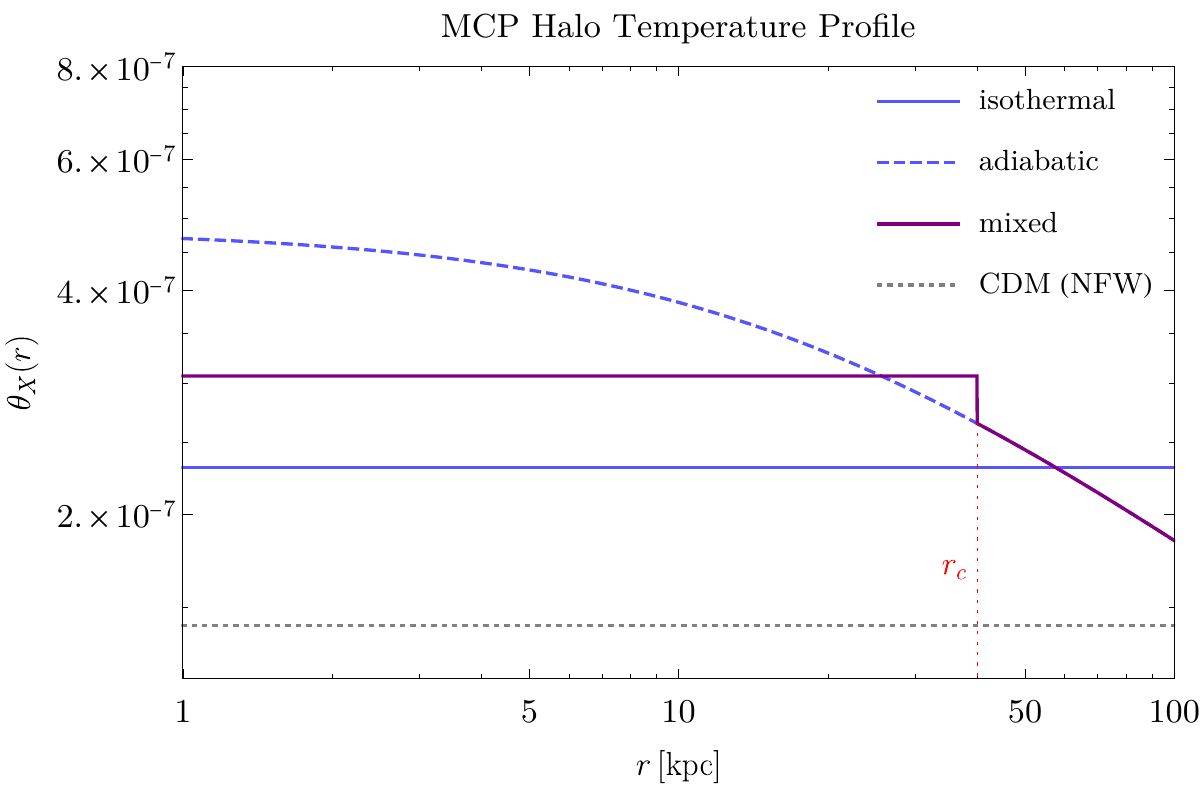}
			\par\end{centering}
		\caption{The mass-density $\rho_\textrm{MCP}$ (left) and the temperature $\theta_X$ (right) profiles of the MCP of the fully-isothermal, fully-adiabatic and ``mixed'' halos. We assume $r_c=\unit[40]{kpc}$ for the ``mixed'' halo.\label{Fig. Mixed MCP Halo Mass-density and Temperature}}
	\end{figure}

	In the intermediate $m_\ell$ mass range this distinction blurs, and we find that only beyond some critical radius $r_c$ heat transfer is inefficient (see e.g. the blue curve in Fig.~\ref{Fig. Conduction Profile}). To estimate the effect of the outer halo, where heat conduction is inefficient, we crudely separate the halo into two --- the isothermal inner halo and the outer regions of the halo that are still in the original adiabatic configuration. The crossover radius is taken to be $r_c$ where the thermalization time coincides with the age of the galaxy. The inner halo temperature and density distribution is solved using the hydrostatic equilibrium Eq.~\eqref{Eq. Single Hydrostatic Equation} as before (see Appendix~\ref{App. Mixed Halo} for more details) and plotted together with the outer adiabatic halo profile in Fig.~\ref{Fig. Mixed MCP Halo Mass-density and Temperature}.

    We will treat the plasma as isothermal if $r_c>\unit[40]{kpc}$, since for $r_c=\unit[40]{kpc}$ the resulting local density and temperature in the mixed-halo coincide with the fully isothermal calculation to within less than $50\%$ error, which corresponds to the right of the dashed purple lines in Fig.~\ref{Fig. m_ell-Q Parameter Space}. We note that this crossover regime coincides with the regime where cooling becomes important (see in Sec.~\ref{Sec. Cooling}), and so requires a full analysis of both effects which is left to a future study. We will therefore treat the halo as isothermal in our parameter range.
	
	\subsection{Cooling}\label{Sec. Cooling}
	
	The MCP plasma is cooled via two main processes, bremsstrahlung emission and inverse Compton scattering with dark CMB photons --- relics from the early universe whose temperature we will take as a free parameter. Bremsstrahlung is the process of $\ell$ particles accelerating due to collisions with $X$ particles and radiating away some of their initial energy. Inverse Compton scatterings are elastic $\ell+\gamma_D\rightarrow\ell+\gamma_D$ collisions, where the $\ell$ particles transfer energy to the dark CMB photons. In both processes the kinetic energy of $\ell$ particles is lost to the dark photons, cooling the MCP plasma. We investigate in what regions of the parameter space MCP cooling is important. 
	
	The rate of energy loss per unit volume due to bremsstrahlung is~\cite{Rosenberg:2017qia}
	\begin{equation}
		P_\textrm{brems}=n_Xn_\ell\frac{16}{3}\sqrt{\frac{2\pi}{3}}\frac{\alpha_D^3}{m_\ell^2}\sqrt{m_\ell T}\bar{g}_\textrm{ff},
	\end{equation}
	where $\bar{g}_\textrm{ff}$ is the thermally averaged free-free Gaunt factor, which encodes quantum corrections to this formula. In our thermalized scenario the dark photon energy is roughly $\omega\approx m_\ell v_\ell^2\sim T$, so the classical limit where $\bar{g}_\textrm{ff}=1$ is sufficiently accurate~\cite{gauntfactorapproximation}. The formula above is computed in the non-relativistic limit, and relativistic corrections are of order $\theta_\ell^{3/2}$~\cite{1980ApJ...238.1026G}. In the parameter space of interest $\theta_\ell\lesssim3\cdot10^{-3}$, and these corrections are subdominant.

    Since the inner regions of the halo are denser than its outer regions, they will lose energy and cool down faster. Heat conduction then will kick in and thermalize the halo,\footnote{We have checked that conduction is always faster than the cooling rate.} and the effects of cooling will be felt throughout the isothermal region of the halo. To neglect its effects on the local MCPs we wish to observe, the energy lost to cooling by the isothermal region of the halo must be small. We therefore can neglect the effects of cooling if the total energy lost within the lifetime of the galaxy is less than some fraction of this energy, chosen here to be $1/3$. 

    The conduction process is effective in the inner region within $r_c>\unit[40]{kpc}$ (see Sec.~\ref{Sec. Thermal Conduction}). The total energy of this region is $E_\textrm{iso}\pqty{r_c}=U_\textrm{iso}\pqty{r_c}+K_\textrm{iso}\pqty{r_c}$ given in Eq.~\eqref{Eq. Explicit Isothermal Energy} in App.~\ref{App. Mixed Halo}. The total energy loss rate of this region of the MCP halo due to bremsstrahlung is then
    \begin{equation}\label{Eq. total brems cooling rate}
        \pqty{\dv{E}{t}}_\textrm{brems}=\int^{r_c} P_\textrm{brems}\dd{V}
        =\frac{1}{m_\ell^3 \pqty{m_X/m_\ell}^{3/2}}\frac{16}{3}\sqrt{\frac{2\pi}{3}}\alpha_D^3\bar{g}_\textrm{ff}\sqrt{\theta_X}\int^{r_c}\rho_\textrm{MCP}^24\pi r^2\dd{r},
    \end{equation}
    and we estimate that within the lifetime of the galaxy a total of $E_\textrm{cool}\approx\pqty{\dv*{E}{t}}_\textrm{brems}t_\textrm{gal}$ was lost. Imposing the condition  $\abs{E_\textrm{cool}/E_\textrm{iso}\pqty{r_c}}\lesssim1/3$ we find that cooling effects on the MCPs can be neglected for masses above
    \begin{equation}\label{Eq: minimal cooling mass}
        m_\ell\gtrsim\unit[1.1]{MeV}\pqty{\frac{10^3}{m_X/m_\ell}}^{1/2}.
    \end{equation}
    We plot the minimal $m_\ell$ values allowed by cooing in Fig.~\ref{Fig. m_ell-Q Parameter Space} as solid blue lines for two values of the mass ratio $m_X/m_\ell=\Bqty{10^3,10^4}$.  
	
	The MCP plasma is also cooled via inverse Compton scattering. Its timescale is controlled by the current dark CMB temperature $T_D^0$ and the redshift $z=2$ at which the Milky Way and its halo were starting to form. The energy loss rate per unit volume by inverse Compton scattering is~\cite{Rosenberg:2017qia}
	\begin{equation}
		P_\textrm{Compton}=\frac{4T_\ell}{m_\ell}\sigma_Tn_\ell\frac{\pi^2}{15}\pqty{T_D^0\pqty{1+z}}^4,
	\end{equation}
    and the total energy loss rate is
    \begin{equation}
        \pqty{\dv{E}{t}}_\textrm{Compton}=\int P_\textrm{Compton}\dd{V}
        =\frac{32\pi^2}{45m_\ell^3}\alpha_D^2\pqty{T_D^0\pqty{1+z}}^4M_\textrm{MCP}\theta_X.
    \end{equation}
    
    The ratio of energy loss rate via bremsstrahlung to inverse Compton scattering is thus
    \begin{equation}
        \frac{\pqty{\dv*{E}{t}}_\textrm{brems}}{\pqty{\dv*{E}{t}}_\textrm{Compton}}\simeq\pqty{\frac{10^4}{m_X/m_\ell}}^{3/2}\pqty{\frac{0.4T_\textrm{CMB}}{T_D^0}}^4.
    \end{equation}
	For $T_D^0\leq 0.4 T_\textrm{CMB}$ more energy is lost to bremsstrahlung than to inverse Compton scatterings for all mass ratios of interest. Larger values of $T_D^0>0.5T_\textrm{CMB}$ are excluded by $\Delta N_\textrm{eff}$ constraints~\cite{Planck:2018vyg}. We thus ignore Compton scattering in the following discussions.

    The dark sector in our scenario is similar to the SM plasma, which we know cools efficiently to form the galactic disk which is the Milky Way. Indeed, for the SM parameters $m_\ell=m_e,m_X=m_p$ we see from Eq.~\eqref{Eq: minimal cooling mass} that cooling is important in the SM. Additionally, cooling due to compton scattering with the CMB photons is significant, while dark compton scattering with the dark CMB is constrained by $N_{\rm eff}$. It is important to note that as the SM plasma cools down to lower temperatures, more cooling channels become available via atomic processes (e.g. recombination, collisional ionization and collisional excitation). In the dark sector we only consider bremsstrahlung which dominates at high temperatures compared with the Rydberg energy~\cite{Rosenberg:2017qia}. As we are only interested in determining whether cooling is important, it's enough to check this at the initial high temperatures. 

    We further note, that these calculations assume the energy lost to the dark photons isn't reabsorbed back into the plasma in photon-particle collisions. We verify this by computing the optical depth of a dark photon travelling from the center of the halo to its edge (in a straight line for simplicity), which is largest for $\gamma_D-\ell$ scatterings,
	\begin{equation}
		\tau_{\gamma_D}=\int_0^{R_\textrm{halo}}\sigma_Tn_\ell\dd{r}=0.03\pqty{\frac{\unit[0.1]{MeV}}{m_\ell}}^3\pqty{\frac{10^3}{m_X/m_\ell}}.
	\end{equation}
	
    When $\tau_{\gamma_D}\ll 1$, the photons escape the galaxy and carry away energy from the plasma, so it is cooled efficiently. Otherwise, the MCP plasma still loses energy via the dark photons, but only from the outer layer of the galaxy, resulting in much more complicated cooling dynamics which we leave for a future study. We determined that reabsorption of dark photons in the halo becomes relevant for masses satisfying $\tau_{\gamma_D}\geq1$. For our two choices of the ratio $m_X/m_\ell=\Bqty{10^3,10^4}$, we find that $\gamma_D$ escape the halo for all $\ell$ masses above $\unit[\Bqty{31,14}]{keV}$, as shown by the dashed blue lines in Fig.~\ref{Fig. m_ell-Q Parameter Space}.
	
	\section{Direct-detection Prospects}\label{Sec. Direct-detection Prospects}
	
	In this section we calculate the experimental signatures of the light MCP $\ell$ in electron-scattering based direct-detection experiments. These experiments detect ionized electrons from a target material on Earth, searching for an excess of events above the expected background rate. The excess electrons were scattered off the target atoms by galactic DM. The incoming DM must surpass a momentum threshold to transfer enough energy to the scattered electron for it to overcome the atomic binding energy. It is typically assumed the DM has virial velocity $\sim10^{-3}$, which limits the experiment's sensitivity to low DM masses. This is not the case in our scenario where the $\ell$'s speed is above the virial velocity, which allows low-mass particles to pass the detection threshold. This is a generic prediction of boosted DM --- direct-detection experiments are able to probe boosted DM in lower mass values than in the standard virialized DM scenario.
	
	The incoming $\ell$ scatters off a bound electron in the detector and excites it, the kinematics of which we solve in Sec.~\ref{Sec. e-l Kinematics}. This scattering is governed by the matrix element, which we calculate in Sec.~\ref{Sec. Matrix Element Approximation}. We then compute the rate of $\ell-e^-$ scattering events expected to be measured in electron-recoil based direct-detection experiments such as XENON10 and SENSEI in Sec.~\ref{Sec. Relativistic Scattering Rate}.
	
	\subsection{Bound Electron-Relativistic MCP Kinematics}\label{Sec. e-l Kinematics}
	
	The light MCP $\ell$ travels through the galaxy and happens to hit the target material in our terrestrial experiment, scattering off an electron of a target atom. The MCP transfers energy and momentum to the scattered electron, and may excite the electron from one energy level to another. We call the initial (final) electron state energy level 1 (2), whose identity will depend on the experiment. To derive the cross-section and event rate, we first consider the kinematics of this process. Derivations of the event rate where the DM is relativistic have been carried out in the literature before, which we review here.
	\begin{figure}
		\begin{centering}
			\begin{tikzpicture}
				\begin{feynman}[large]
					\vertex (a1) at (0,0){\(\ell\)};
					\vertex (a2) at ($(a1) + (4,0)$) {\(\ell\)};
					\vertex (b) at ($(a1) + (2,-1)$);
					\vertex (c) at ($(b) + (0,-1.5)$);
					\vertex (d1) at ($(c) + (-2,-1)$) {$e^{-}$};
					\vertex (d2) at ($(c) + (2,-1)$) {$e^{-}$};
					\diagram*{
						(a1) -- [fermion,momentum'=$p_\ell$] (b) -- [fermion,momentum'=$p'_\ell$] (a2),
						(c) -- [boson, edge label=\(\gamma\),rmomentum'=$q$] (b),
						(d1) -- [fermion,momentum=$k_e$] (c) -- [fermion,momentum=$k'_e$] (d2),
					};
				\end{feynman}
			\end{tikzpicture}
			\par\end{centering}
		\caption{$\ell$ scatters off $e^-$ via $\gamma$ exchange. \label{Fig. electron-MCP Scattering Diagram}}
	\end{figure}
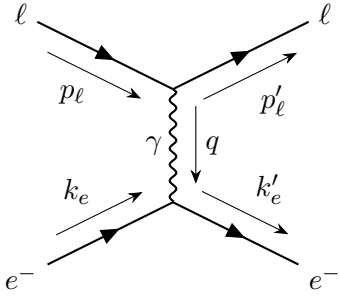
	
	In our model the milli-charged $\ell$ scatters off the bound $e^-$ via photon exchange, as depicted in Fig.~\ref{Fig. electron-MCP Scattering Diagram}. We denote $p_\ell^{\pqty{\prime}}=\pqty{E_\ell^{\pqty{\prime}},\vb{p}_\ell^{\pqty{\prime}}}$ as the incoming (outgoing) $\ell$ momentum, and $\vb*{\beta}^{\pqty{\prime}}_\ell$ its respective velocities (with $\gamma^{\pqty{\prime}}_\ell$ the Lorentz factor). We denote $k_e^{\pqty{\prime}}=\pqty{E_e^{\pqty{\prime}},\vb{k}_e^{\pqty{\prime}}}$ as the electron momentum and $E_{e,1\pqty{2}}$ its total energy minus its mass in level 1 (2). The typical velocity of a bound electron is $v_e\sim Z_\textrm{eff}\alpha/n$ (where $Z_\textrm{eff}$ is the effective charge felt by the electron and $n$ is its principle quantum number), so we  treat the electron non-relativistically. $q$ is the momentum transferred through the photon. In summary the four-momenta are
	\begin{equation}\label{Eq. Specified 4-momenta}
        \begin{aligned}
			p_\ell^{\mu\pqty{\prime}}&=\gamma_\ell^{\pqty{\prime}}m_\ell
			\pqty{1,\vb*{\beta}_\ell^{\pqty{\prime}}},&
			k_e^{\mu\pqty{\prime}}&
			=\pqty{m_e+E_{e,1\pqty{2}},\vb{k}_e^{\pqty{\prime}}},&
			q^{\mu}&=\pqty{\Delta E_{1\rightarrow2},\vb{q}},
		\end{aligned}
	\end{equation}
	where $\Delta E_{1\rightarrow2}=E_{e,2}-E_{e,1}$. They are related by momentum conservation
	\begin{equation}\label{Eq. Photon Momentum Conservation}
		p_\ell-p'_\ell=q=k'_e-k_e.
	\end{equation}
	
	\subsection{Off-shell Matrix Element Approximation}\label{Sec. Matrix Element Approximation}
	
	We turn to the matrix-element calculation using the Lagrangian in Eq.~\eqref{Eq. no mix Lagrangian}. The tree-level amplitude of an electron $e^{-}$ scattering with a particle $\ell$ of charge $-Qe$ corresponding to the diagram in Fig.~\ref{Fig. electron-MCP Scattering Diagram} reads
	\begin{equation}\label{Eq. Amplitude Formula}
		i\mathcal{M}=\bar{u}\pqty{k'_e}\pqty{-ie\gamma^{\mu}}u\pqty{k_e}
		\frac{-i\eta_{\mu\nu}}{q^{2}}\bar{u}\pqty{p'_\ell}\pqty{-ieQ\gamma^{\nu}}u\pqty{p_\ell}.
	\end{equation}
	The MCP is not polarized, so we average over the MCP spin states. We are blind to the scattered electron's spin, so we sum over its spin states. The square of the matrix element in Eq.~\eqref{Eq. Amplitude Formula} is
	\begin{equation}\label{Eq. Spin Averaged Matrix}
		\begin{aligned}
			\frac{1}{4}\sum_{\textrm{spins}}\abs{\mathcal{M}}^{2} & =\frac{e^{4}Q^{2}}{4q^{4}}\textrm{tr}\bqty{\pqty{\slashed{k}'_e+m_{e}}\gamma^{\mu}\pqty{\slashed{k}_e+m_{e}}\gamma^{\nu}}
			\textrm{tr}\bqty{\pqty{\slashed{p}'_\ell+m_{\ell}}\gamma_{\mu}\pqty{\slashed{p}_\ell+m_{\ell}}\gamma_{\nu}}\\
			& =\frac{8e^{4}Q^{2}}{q^{4}}
			\left[\pqty{k'_e\cdot p'_\ell}\pqty{k_e\cdot p_\ell}
			+\pqty{k'_e\cdot p_\ell}\pqty{k_e\cdot p'_\ell}
			-\pqty{k'_e\cdot k_e}m_{\ell}^{2}\right.\\
			&\hphantom{=\frac{8e^{4}Q^{2}}{q^{4}}[\pqty{k'_e\cdot p'_\ell}\pqty{k_e\cdot p_\ell}+\pqty{k'_e\cdot p_\ell}\pqty{k_e\cdot p'_\ell}}
			\left.-\pqty{p'_\ell\cdot p_\ell}m_{e}^{2}+2m_{e}^{2}m_{\ell}^{2}\right].
		\end{aligned}
	\end{equation}
	
	If the electron were free, both particles would be on-shell and the expression above may be further simplified using Mandelstam variables. In direct-detection experiments the electron is bound and therefore off-shell, so we refrain from this simplification. Instead, we use momentum conservation, Eq.~\eqref{Eq. Photon Momentum Conservation}, to substitute $p'_\ell$ and $k_e$ with $q$ in Eq.~\eqref{Eq. Spin Averaged Matrix}:
	\begin{equation}
		\begin{aligned}
			\overline{\abs{\mathcal{M}}^{2}} &
			=\frac{8e^{4}Q^{2}}{q^{4}}\left[\pqty{k'_e\cdot\pqty{p_\ell-q}}
			\pqty{\pqty{k'_e-q}\cdot p_\ell}+\pqty{k'_e\cdot p_\ell}\pqty{\pqty{k'_e-q}\cdot\pqty{p_\ell-q}}\right.\\
			& \hphantom{=\frac{8e^{4}Q^{2}}{q^{4}}\pqty{k'_e\cdot\pqty{p_\ell-q}}}
			\left.-\pqty{k'_e\cdot\pqty{k'_e-q}}m_{\ell}^2-\pqty{\pqty{p_\ell-q}\cdot p_\ell}m_{e}^{2}+2m_{e}^{2}m_{\ell}^{2}\right]\\
			&=\frac{8e^{4}Q^{2}}{q^{4}}\left[2\pqty{k'_ep_\ell}^2-2\pqty{k'_ep_\ell}\pqty{qp_\ell}-2\pqty{k'_eq}\pqty{k'_ep_\ell}+\pqty{k'_eq}\pqty{qp_\ell}+\pqty{k'_ep_\ell}\pqty{q^2}\right.\\
			& \hphantom{=\frac{8e^{4}Q^{2}}{q^{4}}}
			\left.\phantom{2\pqty{k'_ep_\ell}^2-2\pqty{k'_ep_\ell}\pqty{qp_\ell}}-\pqty{k^{\prime2}_e-k'_eq}m_{\ell}^{2}-\pqty{p^2_\ell-qp_\ell}m_{e}^{2}+2m_{e}^{2}m_{\ell}^{2}\right].
		\end{aligned}
	\end{equation}
	The dot products $qp_\ell$, $k'_ep_\ell$ and $k'_eq$ appearing in this expression depend on the angles between the 3-momenta.  We simplify our calculation by removing this dependence as much as possible: since $\ell$ is on-shell we can use Eq.~\eqref{Eq. Photon Momentum Conservation} to show that $q^2=2qp_\ell$.	The electron is non-relativistic and its rest mass dominates the 3-momentum  $\abs{\vb{k}'_e}\ll E'_e$. After these simplifications we find
	\begin{equation}
		\overline{\abs{\mathcal{M}}^{2}} 
		=\frac{4e^4Q^2}{q^4}\bqty{4m_e^2E_\ell^2+m_e^2q^2
			+k'_eq\pqty{2m_\ell^2-4m_eE_\ell+q^2}}.
	\end{equation}
	
	We may disregard the term proportional to $\vb{k}_e'\cdot\vb{q}$ in $k_e'q$ since its contribution to the cross-section vanishes once we integrate over all $\vb{q}$ directions. This is because the detectors considered are insensitive to the directionality of the scattered particles and their atomic form factors are spherically symmetric (see Eqs.~\eqref{Eq. Ionization Scattering Rate} and~\eqref{Eq. Crystal Scattering Rate} in the next section). After removing this term, we plug in the relevant momenta specified in Eq.~\eqref{Eq. Specified 4-momenta} and derive the effective matrix-element needed in our calculations:
	\begin{equation}\label{Eq. Final Matrix Element}
	   \begin{aligned}
	        \overline{\abs{\mathcal{M}}^{2}}_\textrm{eff}
			\approx\frac{4e^4Q^2}{\pqty{\abs{\vb{q}}^2-\Delta E_{1\rightarrow2}^2}^2}
			&\left[\pqty{\pqty{2\gamma_\ell m_\ell -\Delta E_{1\rightarrow2}}^2-\abs{\vb{q}}^2}m_e^2\right.\\
			&\left.+m_e\Delta E_{1\rightarrow2}\pqty{2m_\ell^2+\Delta E_{1\rightarrow2}^2-\abs{\vb{q}}^2}\right]. 
	   \end{aligned}
	\end{equation}
	
	\subsection{Scattering Rate Formulae with Relativistic MCP}\label{Sec. Relativistic Scattering Rate}
	
	Having calculated the matrix-element in Eq.~\eqref{Eq. Final Matrix Element}, we use it to compute the scattering rate of MCPs in electron-recoil based experiments. First we find the cross-section of $e^-+\ell\rightarrow e^-+\ell$ where the electron is bound. The standard cross-section calculation assumes free asymptotic states and needs to be amended when bound electron sates are considered. We follow the derivation presented in Appendix A of~\cite{Essig:2015cda} to account for the effects of the atomic physics, and extend it to incorporate relativistic MCPs. Our results are equivalent to the derivation in~\cite{An:2017ojc}.
	
	Our starting point is the general cross-section formula in Eq. (A.7) of~\cite{Essig:2015cda}:
	\begin{equation}\label{Eq. General Cross-section Formula}
		\sigma v_{1\rightarrow2}=\int\frac{\dd[3]{\vb{q}}}{\pqty{2\pi}^{3}}
		\frac{1}{16E_{\ell}E_{e}E_{\ell}'E_{e}'}2\pi\delta\pqty{E_f-E_i}\overline{\abs{\mathcal{M}}^2}\times\abs{f_{1\rightarrow2}\pqty{\vb{q}}}^{2}.
	\end{equation}
	$E_{i\pqty{f}}$ is the total energy of the initial (final) state of the system. $f_{1\rightarrow2}\pqty{\vb{q}}$ is the \textit{atomic form factor}, i.e. the wave-function overlap between the initial (energy level 1) and final (energy level 2) states of the electron. It encapsulates the effect of the target material on the scattering event.
	
	The electron is non-relativistic, so its energy is $E_e^{\pqty{\prime}}=m_e$, and we keep the Lorentz factors of the MCP's energy $E_\ell^{\pqty{\prime}}=\gamma_\ell^{\pqty{\prime}}m_\ell$. We obtain
	\begin{equation}
		\sigma v_{1\rightarrow2}=\frac{1}{16\pi m_{e}^{2}m_{\ell}^{2}}
		\int\frac{\dd[3]{\vb{q}}}{4\pi}\frac{1}{\gamma_\ell\gamma'_\ell}\delta\pqty{E_f-E_i}
		\overline{\abs{\mathcal{M}}^2}\times\abs{f_{1\rightarrow2}\pqty{\vb{q}}}^{2}.
	\end{equation}
	To calculate the rate of events we average over the MCP velocity and multiply by the local MCP number density
	\begin{equation}\label{Eq. Detection Rate with Delta Function}
	   \begin{aligned}
	    R_{1\rightarrow2}&=n_\ell^{\textrm{loc}}\int \dd[3]{\vb*{\beta}_\ell} g_\ell\pqty{\vb*{\beta}_\ell}\sigma v_{1\rightarrow2}\\
		&=\frac{n_\ell^{\textrm{loc}}}{16\pi m_e^2m_\ell^2}\int\frac{\dd[3]{\vb{q}}}{4\pi}
		\int \frac{\dd[3]{\vb*{\beta}_\ell}}{\gamma_\ell\gamma'_\ell}\delta\pqty{E_f-E_i}
		\overline{\abs{\mathcal{M}}^2}\times\abs{f_{1\rightarrow2}\pqty{\vb{q}}}^{2}
		g_\ell\pqty{\vb*{\beta}_\ell},  
	   \end{aligned}
	\end{equation}
	where $g_{\ell}\pqty{\vb*{\beta}_\ell}$ is $\ell$'s velocity distribution.
	
	We need eliminate the delta function $\delta\pqty{E_f-E_i}$, that is, to impose energy conservation $E_f=E_i$. We make use of the kinematics specified in Sec.~\ref{Sec. e-l Kinematics}: from Eq.~\eqref{Eq. Specified 4-momenta} we find the total energies of the initial and final states to be
	\begin{equation}
		\begin{aligned}
			E_{i} & =E_e+E_\ell=m_{e}+E_{e,1}+\gamma_\ell m_{\ell},\\
			E_{f} & =E_e'+E_\ell'=m_{e}+E_{e,2}+\gamma'_\ell m_{\ell},
		\end{aligned}
	\end{equation}
	which implies the difference is
	\begin{equation}\label{Eq. Initial-final Energy Difference}
		E_{f}-E_{i}=
		\Delta E_{1\rightarrow2}+\pqty{\gamma^{\prime}_\ell-\gamma_\ell}m_\ell.
	\end{equation}
	We extract $\gamma'_\ell$ from 3-momentum conservation,
	\begin{equation}\begin{aligned}
			\gamma'_\ell m_\ell&=E'_\ell=\sqrt{m_\ell^2+\abs{\vb{p}'_\ell}^2}
			=\sqrt{m_\ell^2+\abs{\vb{p}_\ell-\vb{q}}^2}
			=\sqrt{m_\ell^2+\abs{\vb{p}_\ell}^2+\abs{\vb{q}}^2-2\vb{q}\cdot\vb{p}_\ell}\\
			&=\sqrt{\gamma_\ell^2m_\ell^2+\abs{\vb{q}}^2
				-2\abs{\vb{q}}m_\ell\gamma_\ell\beta_\ell\cos\theta},
		\end{aligned}
	\end{equation}
	where $\theta$ is the angle between $\vb{q}$ and $\vb{p}_\ell$. The energy difference in Eq.~\eqref{Eq. Initial-final Energy Difference} is then
	\begin{equation}\label{Eq. Energy Difference with Angles}
		E_{f}-E_{i}
		=\Delta E_{1\rightarrow2}+\sqrt{\gamma_\ell^2m_\ell^2+\abs{\vb{q}}^2
			-2\abs{\vb{q}}m_\ell\gamma_\ell\beta_\ell\cos\theta}-\gamma_\ell m_{\ell}.
	\end{equation}
	
	We find the minimal $\ell$ velocity required to excite the electron by solving the energy conservation equation $E_{i}=E_{f}$ in Eq.~\eqref{Eq. Energy Difference with Angles} when the MCP scatters co-linearly, $\cos\theta=1$. It is 
	\begin{equation}\label{Eq. Minimal Excitation Velocity}
		v_{\min}\pqty{\abs{\vb{q}},\Delta E_{1\rightarrow2}}=\frac{
			4\Delta E_{1\rightarrow2}m_\ell^2\abs{\vb{q}}+
			\sqrt{\pqty{\abs{\vb{q}}^2-\Delta E_{1\rightarrow2}^2}^3\pqty{4m_\ell^2+\abs{\vb{q}}^2-\Delta E_{1\rightarrow2}^2}}}
		{4m_\ell^2\abs{\vb{q}}^2+\pqty{\abs{\vb{q}}^2-\Delta E_{1\rightarrow2}^2}^2}.
	\end{equation}
	In the non-relativistic limit where $m_\ell\gg\abs{\vb{q}}\gg\Delta E_{1\rightarrow2}$ this expression reduces to Eq.~(A.16) in~\cite{Essig:2015cda}.
	
	In our model, the lower integration limit for the transferred 3-momentum is $\abs{\vb{q}}_{\min}=\Delta E_{1\rightarrow2}$, which by Eq.~\eqref{Eq. Minimal Excitation Velocity} corresponds to $v_{\min}\pqty{\abs{\vb{q}}_{\min}}=1$. In contrast, in the standard virialized DM scenario the particle's maximal speed is the escape velocity $v_\textrm{esc}$ and so the excitation velocity must be smaller than it, $v_{\min}\leq v_{\textrm{esc}}$, which imposes the lower integration limit is instead $\abs{\vb{q}}_{\min}^\textrm{vir}\approx \Delta E_{1\rightarrow2}/v_{\textrm{esc}}\sim10^3\Delta E_{1\rightarrow2}$ \cite{Essig:2015cda}. Since in our model we integrate over a larger phase space and smaller momentum transfers $\vb{q}$ (which dominate Rutherford scatterings, see Eq~\eqref{Eq. Final Matrix Element}), the cross-section and rate will be larger and will result in better detector sensitivity compared to those derived in the standard virialized DM scenario.
	
	We eliminate the delta function in Eq.~\eqref{Eq. Detection Rate with Delta Function} with the $\cos\theta$ integral using Eq.~\eqref{Eq. Energy Difference with Angles},
	\begin{equation}\begin{aligned}
	    \int \frac{\dd[3]{\vb*{\beta}_\ell}}{\gamma_\ell\gamma'_\ell}\delta\pqty{E_f-E_i}&=
		\int \frac{\beta_\ell^2\dd{\beta_\ell}\dd{\phi_\beta}}{\gamma_\ell\gamma'_\ell}
		\Theta\pqty{\beta_\ell-v_{\min}}\eval{\abs{\dv{\pqty{E_f-E_i}}{\cos\theta}}^{-1}}_{E_f=E_i}\\
		&=\frac{1}{2\abs{\vb{q}}}\int\frac{\dd[3]{\vb*{\beta}_\ell}}{\gamma_\ell^2\beta_\ell}
		\Theta\pqty{\beta_\ell-v_{\min}},
	\end{aligned}
	\end{equation}
	which results in the scattering rate
	\begin{equation}
		R_{1\rightarrow2}=\frac{n_\ell^{\textrm{loc}}}{16\pi m_e^2m_\ell^2}
		\int\frac{\dd[3]{\vb{q}}}{8\pi}\frac{1}{\abs{\vb{q}}}\abs{f_{1\rightarrow2}\pqty{\vb{q}}}^2
		\int\frac{\dd[3]{\vb*{\beta}_\ell}}{\gamma_\ell^2\beta_\ell}g_\ell\pqty{\vb*{\beta}_\ell}
		\Theta\pqty{\beta_\ell-v_{\min}}\overline{\abs{\mathcal{M}}^2}.
	\end{equation}
	This is consistent with Eq. (A.14) of~\cite{Essig:2015cda} when $\gamma_\ell=1$.
	
	Our choice of $g_\ell\pqty{\vb*{\beta}_\ell}$ is the isotropic \textit{Maxwell-J\"{u}ttner distribution},
	\begin{equation}
		g_{\ell}\pqty{\vb*{\beta}_\ell}\dd[3]{\vb*{\beta}_\ell}
		=\frac{\gamma^{2}_\ell\beta_\ell}{\theta_{\ell}K_{2}\pqty{1/\theta_{\ell}}}
		e^{-\gamma_\ell/\theta_{\ell}}\dd{\gamma_\ell}.
	\end{equation}
	This is the relativistic extension of the Maxwell-Boltzmann distribution usually taken for galactic DM. Here, however, it is not truncated by the DM escape velocity, since $\ell$ particles are still bound to the galaxy even with relativistic velocities. Additionally, we do not account for the relative motion of the Solar System and the MCPs which introduces anisotropy in the distribution function. This is justified because $v_\odot\approx 8\cdot10^{-4}$, way below the typical speed of $\ell$ particles we consider. With this choice of $\ell$'s velocity distribution we find the scattering rate master formula
	\begin{equation}\label{Eq. Scattering Rate Master Formula}
	R_{1\rightarrow2}=\frac{n_{\ell}^{\textrm{loc}}}{16\pi m_e^2m_\ell^2}
	\int\frac{\dd[3]{\vb{q}}}{8\pi}\frac{1}{\abs{\vb{q}}}
	\abs{f_{1\rightarrow2}\pqty{\vb{q}}}^2\frac{1}{\theta_\ell K_2\pqty{1/\theta_\ell}} \int_{\gamma_{\min}}^\infty\overline{\abs{\mathcal{M}}^2}e^{-\gamma_\ell/\theta_{\ell}}\dd{\gamma_\ell},
	\end{equation}
	where $\gamma_{\min}\equiv\gamma\pqty{v_{\min}}$.
	
	Our discussion so far made no detailed assumptions about the experimental apparatus at hand --- we did not specify what the electron's initial and final states are. The effects of the underlying atomic physics of the detector were quantified in the general atomic form factor $f_{1\rightarrow2}\pqty{\vb{q}}$ introduced in Eq.~\eqref{Eq. General Cross-section Formula}. Now we apply our result from Eq.~\eqref{Eq. Scattering Rate Master Formula} for the XENON10 and SENSEI experiments. The derivation is identical to the one in~\cite{Essig:2015cda}, so we shall not repeat it in detail here.
	
	The XENON10 experiment uses a liquid xenon target, searching for electrons ionized from $N_T$ isolated atoms. The atom is assumed spherically symmetric with filled electron shells. Before the scattering event, the electron is bound to the atom with binding energy $E_B$. The electron ionized by the MCP is treated as a continuum of positive-energy bound states, approximated as a free particle at asymptotically large radii. Its 3-momentum is $\vb{k}_e'$ with recoil energy $E_R=\abs{\vb{k}_e'}^2/2m_e$. The ionization rate of an isolated atom is given by
	\begin{equation}\label{Eq. Ionization Scattering Rate}
	\begin{aligned}
	    \dv{R_{\textrm{ion}}}{\ln E_{R}}&=N_{T}\frac{n_{\ell}^{\textrm{loc}}}{16\pi m_e^2
			m_{\ell}^{2}}\frac{1}{8}\int \abs{\vb{q}}\dd{\abs{\vb{q}}}\abs{f_{\textrm{ion}}\pqty{\abs{\vb{k}_e'},\abs{\vb{q}}}}^2\\
			& \hphantom{=N_{T}\frac{n_{\ell}^{\textrm{loc}}}{16\pi m_e^2m_\ell^2}\frac{1}{8}\int}
		\times\frac{1}{\theta_\ell K_2\pqty{1/\theta_\ell}} \int_{\gamma_{\min}}^\infty\overline{\abs{\mathcal{M}}^2}
		e^{-\gamma_\ell/\theta_{\ell}}\dd{\gamma_\ell},
	\end{aligned}
	\end{equation}
	where $\abs{f_{\textrm{ion}}}^2$ is the \textit{ionization form factor} and $\Delta E_{1\rightarrow2}=E_B+E_R$. This is equivalent to Eq.~(A.21) in~\cite{Essig:2015cda}.
	
	The SENSEI experiment uses semiconductor crystals as the target material. The periodic lattice has a continuum of electron energy levels which form a band structure. The valence bands and conduction bands are separated by an energy gap, and as electrons are excited across the bandgap they create mobile electron-hole pairs which can be detected. The MCP hits one of $N_{\textrm{cell}}$ cells in the crystal and deposits $E_e$ energy to an electron which is then excited across the bangap. SENSEI measures the number of electron-hole pairs created by the event, with 1--3 detected pairs set as the threshold. The excitation rate in a semiconductor crystal is given by
	\begin{equation}\label{Eq. Crystal Scattering Rate}
	\begin{aligned}
	    \dv{R_{\textrm{crystal}}}{\ln E_{e}}&=\frac{n_{\ell}^{\textrm{loc}}}{16\pi m_{e}^{2}m_{\ell}^{2}}N_{\textrm{cell}}\alpha_\textrm{QED} m_{e}^{2}
		\int \dd{\ln \abs{\vb{q}}}\frac{E_{e}}{\abs{\vb{q}}}\abs{f_{\textrm{crystal}}\pqty{\abs{\vb{q}},E_{e}}}^2\\
		&\hphantom{=\frac{n_{\ell}^{\textrm{loc}}}{16\pi m_{e}^{2}m_{\ell}^{2}}N_{\textrm{cell}}\alpha_\textrm{QED} m_{e}^{2}
		\int}
		\times\frac{1}{\theta_\ell K_2\pqty{1/\theta_\ell}} \int_{\gamma_{\min}}^\infty\overline{\abs{\mathcal{M}}^2}
		e^{-\gamma_\ell/\theta_{\ell}}\dd{\gamma_\ell},
	\end{aligned}
	\end{equation}
	where $\abs{f_{\textrm{crystal}}}^2$ is the \textit{crystal form factor} and $\Delta E_{1\rightarrow2}=E_{e}$. This is equivalent to Eq.~(A.32) in~\cite{Essig:2015cda}.
	
	\section{Detector Reach and Predictions}
	
\begin{figure}
		\begin{centering}
			\includegraphics[width=0.5\columnwidth]{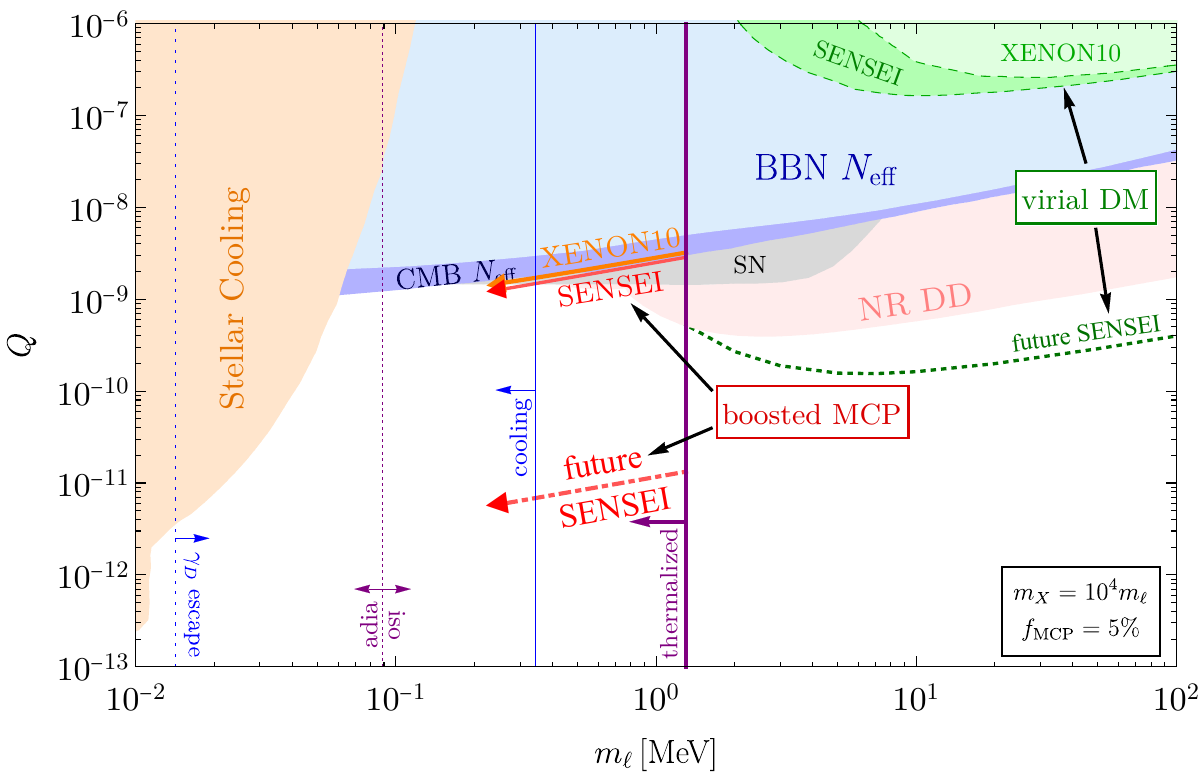}\includegraphics[width=0.5\columnwidth]{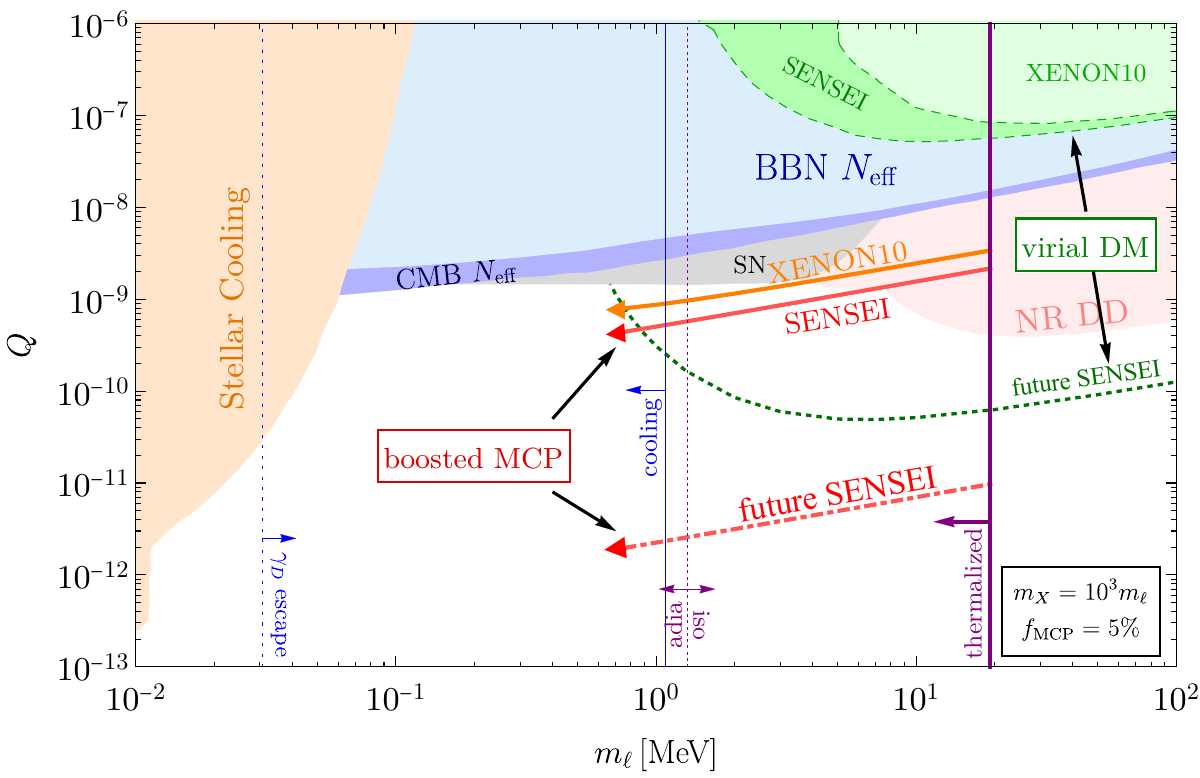}
			\par\end{centering}
		\caption{The parameter space of our model in the $m_\ell-Q$ plane. The potential detector reach of XENON10 in orange and of SENSEI in red (dash-dotted is a future prediction) are for the boosted MCP. The purple and blue vertical lines show the crossover parameter regions for the halo behavior derived in Sec.~\ref{Sec. Thermodynamic Considerations}. Additional constraints are from stellar cooling (light orange) and BBN/CMB $N_{\textrm{eff}}$ (light blue and blue)~\cite{Vogel:2013raa}, SN1987A (gray)~\cite{Chang:2018rso}, XENON10 and SENSEI constraints on virial milli-charged DM, assuming the same $n_\ell$ as in the boosted MCP (light green and green)~\cite{Essig:2017kqs,SENSEI:2020dpa}, the future projection of SENSEI in the virial case (dotted green)~\cite{Essig:2015cda} and nuclear-recoil direct-detection constraints on the $X$ partner (pink)~\cite{Gelmini:2018ogy,Kahlhoefer:2017ddj}. See the text for more details.
		\label{Fig. m_ell-Q Parameter Space}}	
	\end{figure}

	In this section we present the results of our analysis of the direct-detection prospects for our boosted MCP. For this, we compute the expected detection rate of the milli-charged $\ell$ particle in these experiments using Eqs.~\eqref{Eq. Ionization Scattering Rate} and~\eqref{Eq. Crystal Scattering Rate} following the methods outlined in~\cite{Essig:2017kqs,Essig:2015cda,SENSEI:2020dpa}. The projected future reach of SENSEI with the single-pair threshold is shown in the dash-dotted red curve in Fig.~\ref{Fig. m_ell-Q Parameter Space}, assuming a MCP detection rate of $3.6$ signal events in $\unit[1]{kg-yr}$ (the optimistic benchmark value used in~\cite{Essig:2015cda}). We see that a large part of the parameter space accessible by the future runs of SENSEI is not excluded by the current data of XENON10~\cite{Essig:2017kqs} and SENSEI~\cite{SENSEI:2020dpa} (3-pair threshold\footnote{Current data from the 1- and 2-pair thresholds is dominated by background events, and the curves in the parameter space corresponding to these thresholds lie above the 3-pair solid red curve.}), which probes the region above the solid orange and red curves, respectively. Our analysis of the MCP halo dynamics is simplified; we did not take into account dissipation, different plasma temperatures, the effects of interstellar magnetic fields or plasma instabilites. We therefore refrain from interpreting the solid orange and red curves as direct bounds, pending a more detailed analysis. In green we show the constraints for a standard virial milli-charged DM for the same experiments~\cite{Essig:2017kqs,SENSEI:2020dpa,Essig:2015cda}, assuming the same $\ell$ number-density as in our model, for comparison. As expected, the direct-detection experiments are able to probe boosted milli-charged DM in a significantly lighter mass range and with better sensitivity compared with the standard virial case.

	We move to analyze the constraints on our scenario. In the pink region labeled ``NR DD'' we plot the nuclear-recoil direct-detection bounds on the heavier MCP $X$. This species is non-relativistic and is subject to the usual constraints set by these experiments. We translate the bounds from the $m_X-Q$ plane to the $m_\ell-Q$ plane, keeping the mass ratio fixed at $m_X/m_\ell=10^4$ and $m_X/m_\ell=10^3$, respectively. The constraints on $X$ were obtained by XENON1T~\cite{XENON:2019gfn}, PandaX-II~\cite{PandaX-II:2017hlx}, LUX~\cite{LUX:2018akb}, PICO~\cite{PICO:2019vsc}, CRESST~\cite{CRESST:2019jnq} and  CDMSlite~\cite{SuperCDMS:2018gro}, derived in~\cite{Gelmini:2018ogy,Kahlhoefer:2017ddj} for light mediator exchange. We also include indirect constraints from BBN/CMB. Charged DM and massless dark photons contribute to the radiation density of the Universe at matter-radiation equality to which the CMB spectrum is sensitive, and the effect is usually parameterized by the effective number of neutrinos $N_\textrm{eff}$. The additional radiation also changes the relic abundance of primordial elements generated by BBN. The bounds on $N_\textrm{eff}$ from CMB and BBN are translated into bounds on the MCP mass and charge in~\cite{Vogel:2013raa}. These are shown in blue and light blue contours in Fig~\ref{Fig. m_ell-Q Parameter Space}. Inside stars MCPs are produced in pairs by plasmon decay and contribute to the total energy loss of the star. The MCP cooling channel modifies stellar evolution and shortens the lifetime of stars, thereby changing their expected population. The observed population of red giants, horizontal-branch stars and white dwarfs therefore constrain these additional channels of stellar cooling~\cite{Vogel:2013raa}. A similar argument can also be applied to the supernova SN1987A: the expected number of neutrinos would decrease if the proto-neutron star is cooled by more particles, which can be compared to the observed flux of the supernova~\cite{Chang:2018rso}. Recently, however, questions were raised with regards to how robust these bounds are~\cite{Bar:2019ifz}. Stellar cooling bounds are shown in light orange and bounds from SN1987A are shown in gray in Fig~\ref{Fig. m_ell-Q Parameter Space}.
	
	We now consider the regions of parameter space wherein our assumptions regarding the MCP's distribution and speed are justified. The solid purple line indicates the allowed region in which the $X$ and $\ell$ plasmas are thermalized, as explained in Sec.~\ref{Sec. l-X Thermalization}. To the right of the purple line the two plasmas have different temperatures and transfer heat slowly. It is plausible that close to the thermalization bound the $\ell$ plasma is heated enough to boost the particles as in the thermalized scenario we considered, but we leave the full analysis for future work. The solid blue line indicates the boundary of the region in which the bremsstrahlung emission rate is sufficiently high and cools the MCP plasma (Sec.~\ref{Sec. Cooling}), allowing for a formation of a dark disk within the galaxy, resulting in lower MCP velocities $v_X\sim10^{-4}$~\cite{Fan:2013yva,Fan:2013tia,Chacko:2021vin,McCullough:2013jma,Agrawal:2017rvu}. In the cooled MCP region further investigation of the MCP halo dynamics is needed to draw definitive conclusions about our model. We thus refrain from plotting the detection reach in this region, although we do not rule out the possibility this region may still be within the reach of SENSEI. The dashed purple line indicates the crossover between the isothermal and adiabatic halo as dictated by the thermal conduction timescale (see Sec.~\ref{Sec. Thermal Conduction}). As we can see, the halo can be treated as isothermal wherever cooling is negligible. Beyond the dashed blue line the dark photon's optical depth is greater than one, so dark photons emitted in cooling processes are reabsorbed into the plasma, further complicating the dissipative dynamics. Additionally, throughout the paper we assumed the MCPs sustain a spherical halo. It is conceivable that due to gravothermal collapse the halo had already collapsed within the lifetime of our galaxy. In Appendix~\ref{App. Gravothermal Collapse} we argue why the MCP halo will not undergo gravothermal collapse and is stable in our parameter region of interest.
	
	Finally, we have presented the results for two values of $m_X/m_{\ell}$. As $m_{X}/m_{\ell}$ increases, direct-detection experiments are able to probe smaller $m_\ell$ masses in the parameter space. This is because the higher temperature $T/m_\ell\propto m_{X}/m_{\ell}$ increases the average velocity of $\ell$ particles. With higher speeds, lighter MCPs are able to overcome the momentum threshold required to excite the bound electron in the detector (Sec.~\ref{Sec. Direct-detection Prospects}). For even higher values of $m_X/m_\ell$, SENSEI would only be able to probe our model in an already excluded region due to stellar cooling bounds, while mass values above $10^{-2}$--$\unit[10^{-1}]{MeV}$ would be beyond the thermalization limit, where the effects of heat transfer need to be understood in order to give meaningful predictions about our model.
	
	\section{Conclusions}
	
	In this work we explored a two-species MCP DM sub-component where a large mass hierarchy boosts the lighter particles due to thermalization with their heavy partners. The dark charge of the halo prevents the light MCP from escaping it despite its high velocity. We evaluated the MCP density in thermal and hydrostatic equilibria in the presence of a dominant neutral CDM component. We found the relevant parameter space where the MCP components are thermalized and dissipative dynamics can be neglected, and determined that the MCP halo can be treated as isothermal in our parameter range. We then analyzed the detection rate of the lighter species in direct-detection experiments XENON10 and SENSEI, concluding that the boosted MCP can be probed in lower masses as anticipated.
	
	The light MCP mass range in our model is within $\unit[100]{keV}$ to $\unit[10]{MeV}$. The upper mass limit of our analysis is set by the thermalization requirement, above which the two MCP plasmas are not in thermal equilibrium. Current experimental constraints access parts of the available parameter space, while large parts remain unexplored and could be probed by future runs of SENSEI. Much of our lower mass parameter space is subject to significant cooling, which should be considered in a future study. A departure from the isothermal halo is also expected in this region. 
	
	\acknowledgments{We thank Tomer Volansky, Nadav Joseph Outmezguine, Omer Katz and Eric Kuflik for useful discussions. We give special thanks to Itay Mimouni Bloch for his review and helpful comments about the direct-detection prospects. MG is supported in part by Israel Science Foundation under Grant No. 1302/19. MG is also supported in part by the US-Israeli BSF grant 2018236.}
	
	\begin{appendices}

	\section{Analytic Results for the MCP Galactic Distribution}\label{App. MCP Halo Analytic Results}
	
	\subsection{Isothermal MCP Halo}
	
	The CDM makes up the majority of the DM in the halo and we assumed it has the standard NFW profile. Its mass density, mass profile and induced gravitational potential are given by
	\begin{equation}\begin{aligned}
			\rho_{\textrm{NFW}}\pqty{r}&=\rho_{H}\frac{1}{x\pqty{1+x}^2},\\
			M_{\textrm{CDM}}\pqty{r}&=\int\rho_{\textrm{NFW}}4\pi r^{2}\dd{r}
			=4\pi\rho_{H}R_{H}^{3}\bqty{\ln(1+x)-\frac{x}{1+x}},\\
			\Phi\pqty{r}&=\int\frac{GM_{\textrm{CDM}}}{r^{2}}\dd{r}
			=-4\pi G\rho_{H}R_{H}^{2}\frac{\ln(1+x)}{x},
		\end{aligned}
	\end{equation}
	where $x\equiv r/R_H$, with the density scale and scale radius taken to be $\rho_H=1.4\times10^7M_\odot\unit{kpc^{-3}}$ and $R_H=\unit[16.1]{kpc}$ in the Milky Way~\cite{Nesti:2013uwa}. Using the virial theorem we can find the virial temperature of the CDM in Eq.~\eqref{Eq. Virial CDM Temperature} to be
	\begin{equation}
    \begin{aligned}
		\theta_{\textrm{CDM}}&=
		\frac{\displaystyle{\int_0^{R_\textrm{halo}}} GM_{\textrm{CDM}}\rho_{\textrm{NFW}}4\pi r\dd{r}}
		{3M_{\textrm{CDM}}\pqty{R_\textrm{halo}}}\\
		&=\frac{4\pi G\rho_{H}R_{H}^{2}}{3}
		\frac{x_{\max}\pqty{2+x_{\max}}-2\pqty{1+x_{\max}}\ln(1+x_{\max})}{\pqty{1+x_{\max}}
		\bqty{-x_{\max}+\pqty{1+x_{\max}}\ln(1+x_{\max})}}\simeq1.4\times10^{-7},
	\end{aligned}
	\end{equation}
	where $x_{\max}\equiv R_\textrm{halo}/R_H$.
	
	The isothermal MCP halo distribution is found by solving the hydrostatic equilibrium condition in Eq.~\eqref{Eq. Single Hydrostatic Equation} along with the equation-of-state $P_i=n_iT$ and is given by: 
	\begin{equation}\label{Eq. Explicit Isothermal Mass-density Profile}
		\rho_{\textrm{MCP}}^{\textrm{iso}}\pqty{r}
		=\rho_{\textrm{MCP}}^{\textrm{iso}}\pqty{0}
		\exp\bqty{\frac{\Phi\pqty{0}-\Phi\pqty{r}}{2\theta_{X}^\textrm{iso}}}
		=\rho_{\textrm{MCP}}^{\textrm{iso}}\pqty{0}
		\exp\bqty{\frac{2\pi G\rho_H R_H^2}{\theta_{X}^\textrm{iso}}\pqty{\frac{\ln(1+x)}{x}-1}}.
	\end{equation}
	The central density is found by requiring the condition on the total mass of the MCP halo $M_\textrm{MCP}=f_\textrm{MCP} M_\textrm{CDM}$, where $M_\textrm{CDM}$ is the total mass of the CDM halo and $f_\textrm{MCP}$ is taken as a free parameter. Consequently
	\begin{equation}
		\rho_{\textrm{MCP}}^{\textrm{iso}}\pqty{0}=
		\frac{M_{\textrm{MCP}}}{{\displaystyle\int}
			\frac{\rho_{\textrm{MCP}}^{\textrm{iso}}\pqty{r}}
			{\rho_{\textrm{MCP}}^{\textrm{iso}}\pqty{0}}4\pi r^{2}\dd{r}}
		=\frac{f_{\textrm{MCP}}\times\rho_{H}\bqty{\ln(1+x_{\max})-\frac{x_{\max}}{1+x_{\max}}}}
		{{\displaystyle\int}\exp[\frac{2G\rho_{H}R_{H}^{2}}{\theta_{X}^\textrm{iso}}
			\pqty{\frac{\ln(1+x)}{x}-1}]x^{2}\dd{x}}.
	\end{equation}
	
	The isothermal temperature $\theta_{X}^\textrm{iso}$ is fixed by the requirement of energy conservation in  Eq.~\eqref{Eq. Isothermal Halo Energy Conservation}. $E_\textrm{vir}$ is given in Eq.~\eqref{Eq. Virial CDM Energy}. The potential and kinetic energies of the isothermal halo were found in Eqs.~\eqref{Eq. Isothermal Potential Energy} and~\eqref{Eq. Isothermal Kinetic Energy}, and are computed to be
	\begin{equation}\label{Eq. Explicit Isothermal Energy}\begin{aligned}
		U_\textrm{iso}&=\int\Phi\rho_{\textrm{MCP}}^{\textrm{iso}}4\pi r^{2}\dd{r}\\
		&=-16\pi^{2}G\rho_{H}\rho_{\textrm{MCP}}^{\textrm{iso}}\pqty{0}R_{H}^{5}
		\int\frac{\ln(1+x)}{x}\exp[\frac{2\pi G\rho_{H}R_{H}^{2}}{\theta_{X}^\textrm{iso}}\pqty{\frac{\ln(1+x)}{x}-1}]x^{2}\dd{x},\\
		K_{\textrm{iso}}&=3\theta_{X}^\textrm{iso}M_{\textrm{MCP}}
		=12\pi\theta_{X}^\textrm{iso}f_\textrm{MCP}\rho_{H}R_{H}^{3}\bqty{\ln(1+x_{\max})-\frac{x_{\max}}{1+x_{\max}}}.
		\end{aligned}
	\end{equation}
	By numerically solving Eq.~\eqref{Eq. Isothermal Halo Energy Conservation} we find $\theta_{X}^\textrm{iso}\simeq2.3\times10^{-7}$.
	
	\subsection{Adiabatic MCP Halo}
	
	If heat transfer is inefficient the MCP halo develops a temperature gradient, and is instead adiabatically isolated. For ideal gases the polytropic process equation is $P_i=C_i \rho_i^\gamma$ with adiabatic index $\gamma$ and some constant $C_i$, and combined with the ideal gas law $P_i=\rho_i\theta_i$ this gives the equivalent relation $\theta_i=C_i\rho_i^{\gamma-1}$. The gas constituents are particles ($X$ or $\ell$) without internal degrees of freedom or ``monoatomic'' where $\gamma=5/3$. As in the isothermal regime in Sec.~\ref{Sec. Galactic MCP Distribution} the equal number densities ($n_X=n_\ell$) and temperatures ($T_X=T_\ell$ locally) imply $P_X+P_\ell=2P_X\simeq2C_X\rho_\textrm{MCP}^\gamma$. The adiabatic profile is then given by solving Eq.~\eqref{Eq. Single Hydrostatic Equation}, from which we find the temperature profile to be
	\begin{equation}\label{Eq. Explicit Adiabatic Temperature Profile}\begin{aligned}
		2C_{X}\frac{1}{\rho_{\textrm{MCP}}^{\textrm{adia}}}
		\dv{\pqty{\rho_{\textrm{MCP}}^{\textrm{adia}}}^{\gamma}}{r}&
		=2\frac{\gamma}{\gamma-1}\dv{\theta_{X}^{\textrm{adia}}}{r}=-\dv{\Phi}{r}\\
		\implies\theta_{X}^{\textrm{adia}}\left(r\right)&
		=\theta_{X}^\textrm{adia}\pqty{0}+\frac{\gamma-1}{2\gamma}\bqty{\Phi\pqty{0}-\Phi\pqty{r}}\\
		&=\theta_{X}^\textrm{adia}\pqty{0}
		+2\pi G\rho_{H}R_{H}^{2}\frac{\gamma-1}{\gamma}\pqty{\frac{\ln(1+x)}{x}-1},
		\end{aligned}
	\end{equation}
	and the mass density profile is
	\begin{equation}\label{Eq. Explicit Adiabatic Mass-density Profile}
		\rho_{\textrm{MCP}}^{\textrm{adia}}\pqty{r}
		=\bqty{C_X\theta_{X}^{\textrm{adia}}\pqty{r}}^{\frac{1}{\gamma-1}}
		=\hat{C}_{X}\bqty{\theta_{X}^\textrm{adia}\pqty{0}
			+2\pi G\rho_{H}R_{H}^{2}\frac{\gamma-1}{\gamma}\pqty{\frac{\ln(1+x)}{x}-1}}
		^{\frac{1}{\gamma-1}},
	\end{equation}
	where $\hat{C}_{X}\equiv C_{X}^{1/\pqty{1-\gamma}}$. This constant is set by the requirement $M_\textrm{MCP}=f_\textrm{MCP} M_\textrm{CDM}$,
	\begin{equation}
		\hat{C}_X=
		\frac{M_{\textrm{MCP}}}{{\displaystyle\int}
			\frac{\rho_{\textrm{MCP}}^{\textrm{adia}}\pqty{r}}{\hat{C}_X}4\pi r^{2}\dd{r}}
		=\frac{f_{\textrm{MCP}}\times\rho_{H}\bqty{\ln(1+x_{\max})-\frac{x_{\max}}{1+x_{\max}}}}
		{{\displaystyle\int}\bqty{\theta_{X}^\textrm{adia}\pqty{0}
				+2\pi G\rho_{H}R_{H}^{2}\frac{\gamma-1}{\gamma}\pqty{\frac{\ln(1+x)}{x}-1}}
			^{\frac{1}{\gamma-1}}x^{2}\dd{x}}.
	\end{equation}
	
    Similarly to the previous isothermal case, collisions between the MCPs only serve to redistribute the energy but leave the total energy unchanged, so the adiabatic halo has the same total energy $E_\textrm{vir}$ as in Eq.~\eqref{Eq. Virial CDM Energy}. The total energy of the adiabatic halo is the sum of its potential and kinetic energies:
	\begin{equation}\label{Eq. Explicit Adiabatic Energy}\begin{aligned}
		U_{\textrm{adia}}&=\int\Phi\rho_{\textrm{MCP}}^{\textrm{adia}}4\pi r^{2}\dd{r}
		=\hat{C}_{X}\int\Phi\bqty{\theta_{X}^{\textrm{adia}}}^{\frac{1}{\gamma-1}}4\pi r^{2}\dd{r}\\
		&=-16\pi^{2}G\rho_{H}R_{H}^{5}\hat{C}_{X}\int\frac{\ln(1+x)}{x}
		\bqty{\theta_{X}^\textrm{adia}\pqty{0}
			+2\pi G\rho_{H}R_{H}^{2}\frac{\gamma-1}{\gamma}\pqty{\frac{\ln(1+x)}{x}-1}}
		^{\frac{1}{\gamma-1}}x^{2}\dd{x},\\
		K_{\textrm{adia}}&
		=3\int\rho_{\textrm{MCP}}^{\textrm{adia}}\theta_{X}^\textrm{adia}\dd{V}
		=3\hat{C}_{X}\int\bqty{\theta_{X}^{\textrm{adia}}}^{\frac{\gamma}{\gamma-1}}4\pi r^{2}dr\\
		&=12\pi\hat{C}_{X}R_{H}^{3}\int\bqty{\theta_{X}^\textrm{adia}\pqty{0}
			+2\pi G\rho_{H}R_{H}^{2}\frac{\gamma-1}{\gamma}\pqty{\frac{\ln(1+x)}{x}-1}}
		^{\frac{\gamma}{\gamma-1}}x^{2}\dd{x}.
		\end{aligned}
	\end{equation}
	The central temperature $\theta_X^\textrm{adia}\pqty{0}$ is fixed by the energy conservation requirement
	\begin{equation}
		E_{\textrm{adia}}=K_{\textrm{adia}}+U_{\textrm{adia}}=E_{\textrm{vir}},
	\end{equation}
	which is found to be $\theta_{X}^\textrm{adia}\pqty{0}\simeq4.8\times10^{-7}$.
	
	\subsection{Mixed MCP Halo}\label{App. Mixed Halo}
	
	In Sec.~\ref{Sec. Thermal Conduction} we discussed an intermediate $m_\ell$ mass range in which heat transfer is efficient in the halo only below some critical radius $r_c$. We crudely estimate the behavior of the MCP halo in this scenario as isothermal for all $r<r_c$ and adiabatic for $r>r_c$,
	\begin{equation}\begin{aligned}
		\rho_\textrm{MCP}^\textrm{mixed}\pqty{r}&=\begin{cases}
			\rho^\textrm{iso}\pqty{r} & r<r_c\\
			\rho^\textrm{adia}\pqty{r} & r>r_c
		\end{cases},&
	\theta_X^\textrm{mixed}\pqty{r}&=\begin{cases}
		\theta^\textrm{iso} & r<r_c\\
		\theta_X^\textrm{adia}\pqty{r} & r>r_c
	\end{cases},
		\end{aligned}
	\end{equation}
	with the forms of $\rho^\textrm{iso}$, $\rho^\textrm{adia}$ and $\theta_X^\textrm{adia}$ given in Eqs.~\eqref{Eq. Explicit Isothermal Mass-density Profile},~\eqref{Eq. Explicit Adiabatic Mass-density Profile} and~\eqref{Eq. Explicit Adiabatic Temperature Profile} respectively. Since the halo's original configuration was purely adiabatic, we identify the region $r>r_c$ with the original adiabatic solution found in the preceding section, so the integration constants $\hat{C}_X$ and $\theta_{X}^\textrm{adia}\pqty{0}$ are the same as before.
	
	The remaining integration constants of the isothermal region, the central density $\rho^\textrm{iso}\pqty{0}$ and the isothermal temperature $\theta^\textrm{iso}$, are found by imposing the same constraints as in the previous two sections: (1) the total MCP mass is fixed,
	\begin{equation}
		f_\textrm{MCP}M_\textrm{CDM}=M_\textrm{MCP}
		=\int_0^{r_c}\rho^\textrm{iso}\pqty{r}4\pi r^2\dd{r}+\int_{r_c}^{R_\textrm{halo}}\rho^\textrm{adia}\pqty{r}4\pi r^2\dd{r},
	\end{equation}
	and (2) the total energy of the MCP halo is conserved and given by $E_\textrm{vir}$ in Eq.~\eqref{Eq. Virial CDM Energy},
	\begin{equation}
	\begin{aligned}
	   	E_\textrm{vir}&=E_\textrm{iso}\pqty{0,r_c}+E_\textrm{adia}\pqty{r_c,R_\textrm{halo}}\\
	   	&=U_\textrm{iso}\pqty{0,r_c}+U_\textrm{adia}\pqty{r_c,R_\textrm{halo}}
		+K_\textrm{iso}\pqty{0,r_c}+K_\textrm{adia}\pqty{r_c,R_\textrm{halo}}, 
	\end{aligned}
	\end{equation}
	where the potential and kinetic energies of the halos are given in Eqs.~\eqref{Eq. Explicit Isothermal Energy} and~\eqref{Eq. Explicit Adiabatic Energy} and are understood to be computed only within the region indicated by the parenthesis.
	
	The isothermal, adiabatic and ``mixed'' MCP halo mass density and temperature profiles are shown in Fig.~\ref{Fig. Mixed MCP Halo Mass-density and Temperature}.

	\section{Thermalization and Conduction Times of the MCP Halo}\label{App. Thermodynamic Regime Plots}

    In Sec.~\ref{Sec. Thermodynamic Considerations} we examined some of the thermodynamic properties of the MCP halo, in particular its thermalization time $t_\textrm{th}$ and its conduction time $t_\textrm{cond}$. Here we plot these timescales as a function of the halo radius for a few typical $m_\ell$ masses in our parameter space. For comparison, on all the plots we also show the age of the galaxy $t_\textrm{gal}$ in blue.
	
\begin{figure}
		\begin{centering}
			\includegraphics[width=0.5\columnwidth]{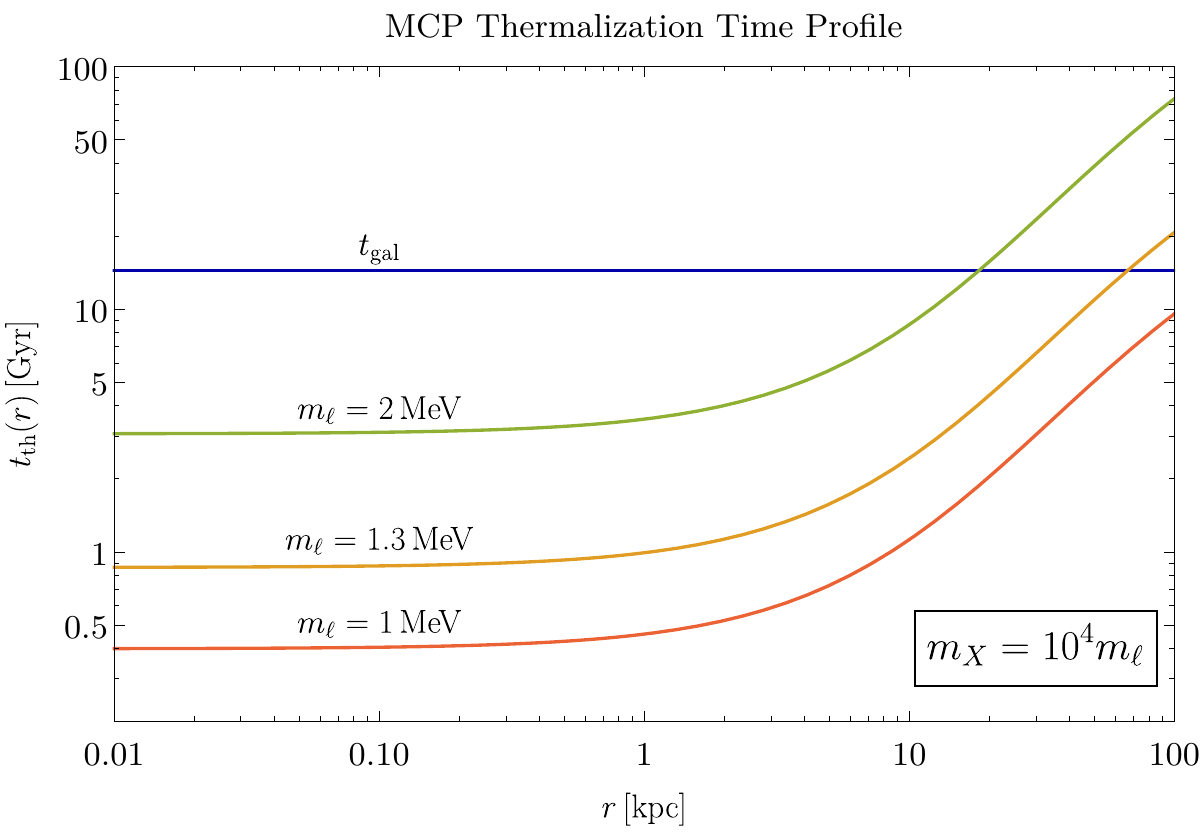}\includegraphics[width=0.5\columnwidth]{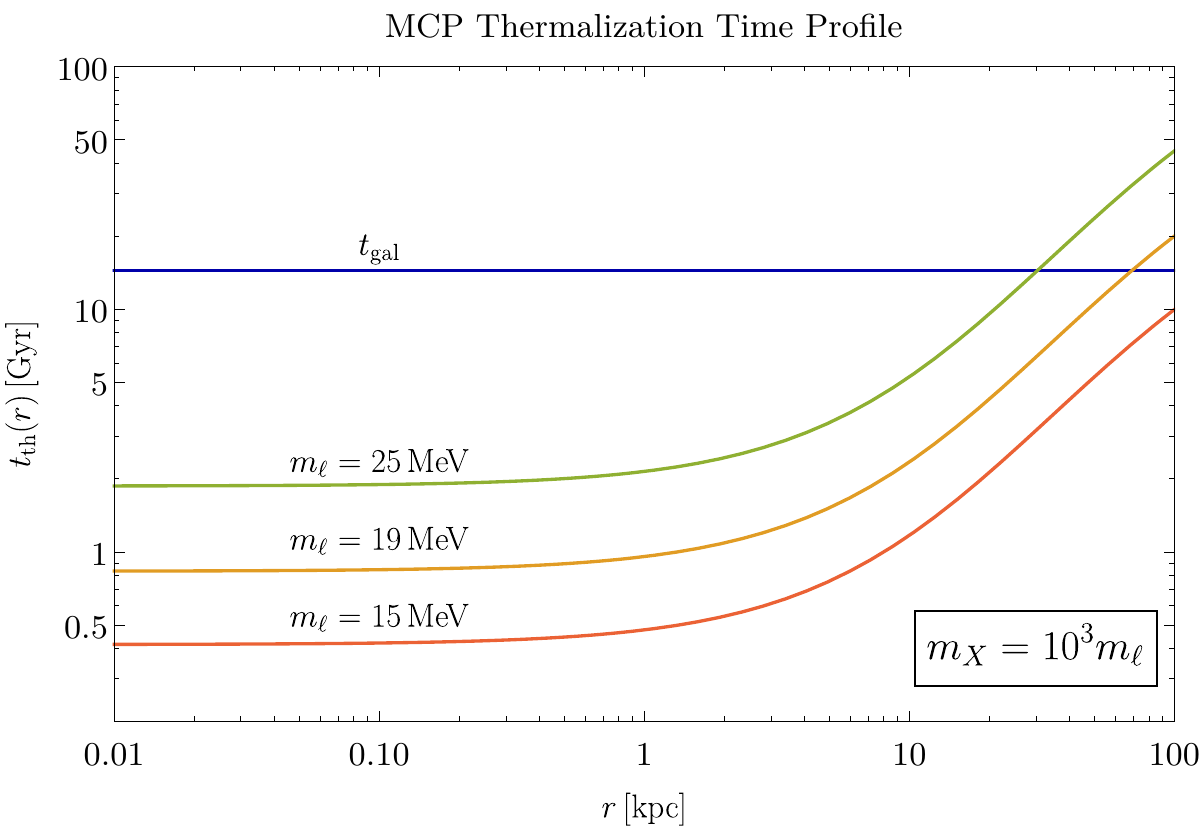}
			\par\end{centering}
		\caption{The thermalization time of the $X$ and $\ell$ plasmas as a function of the radius, with $t_\textrm{gal}$ for comparison. $t_\textrm{th}$ was computed in Sec.~\ref{Sec. l-X Thermalization} using the isothermal halo density in Eq.~\eqref{Eq. Explicit Isothermal Mass-density Profile}. \label{Fig. Thermalization Timescale Profiles}}
	\end{figure}

	The first plots (Fig.~\ref{Fig. Thermalization Timescale Profiles}) compare the thermalization timescale $t_\textrm{th}$ with the age of the galaxy $t_\textrm{gal}$ as a function of the halo radius. The condition for thermalization is conservatively chosen to be $t^{\textrm{avg}}_\textrm{th}\lesssim t_\textrm{gal}$, where $t^{\textrm{avg}}_\textrm{th}$ is calculated using the average number-density of the entire halo. With this condition, the orange curve shows the threshold value of the mass for which this average thermalization condition is saturated. 
	
\begin{figure}
		\begin{centering}
			\includegraphics[width=0.5\columnwidth]{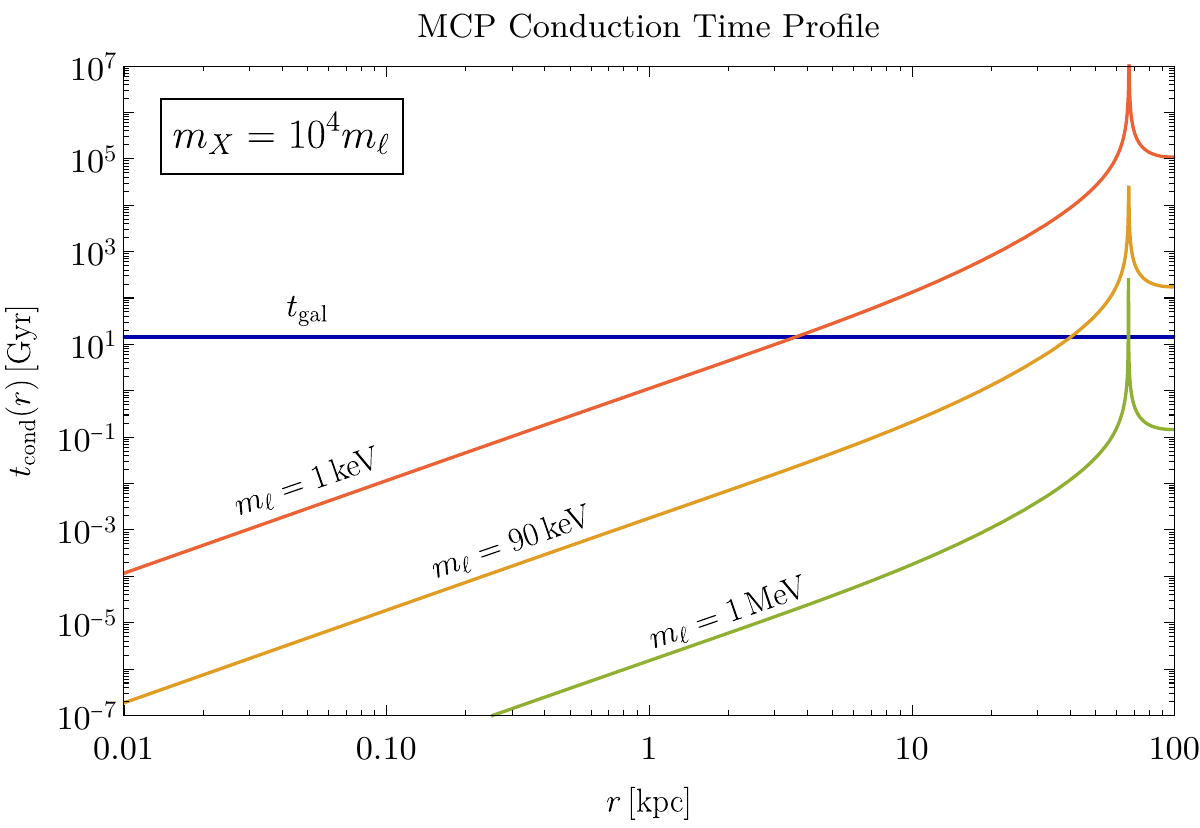}\includegraphics[width=0.5\columnwidth]{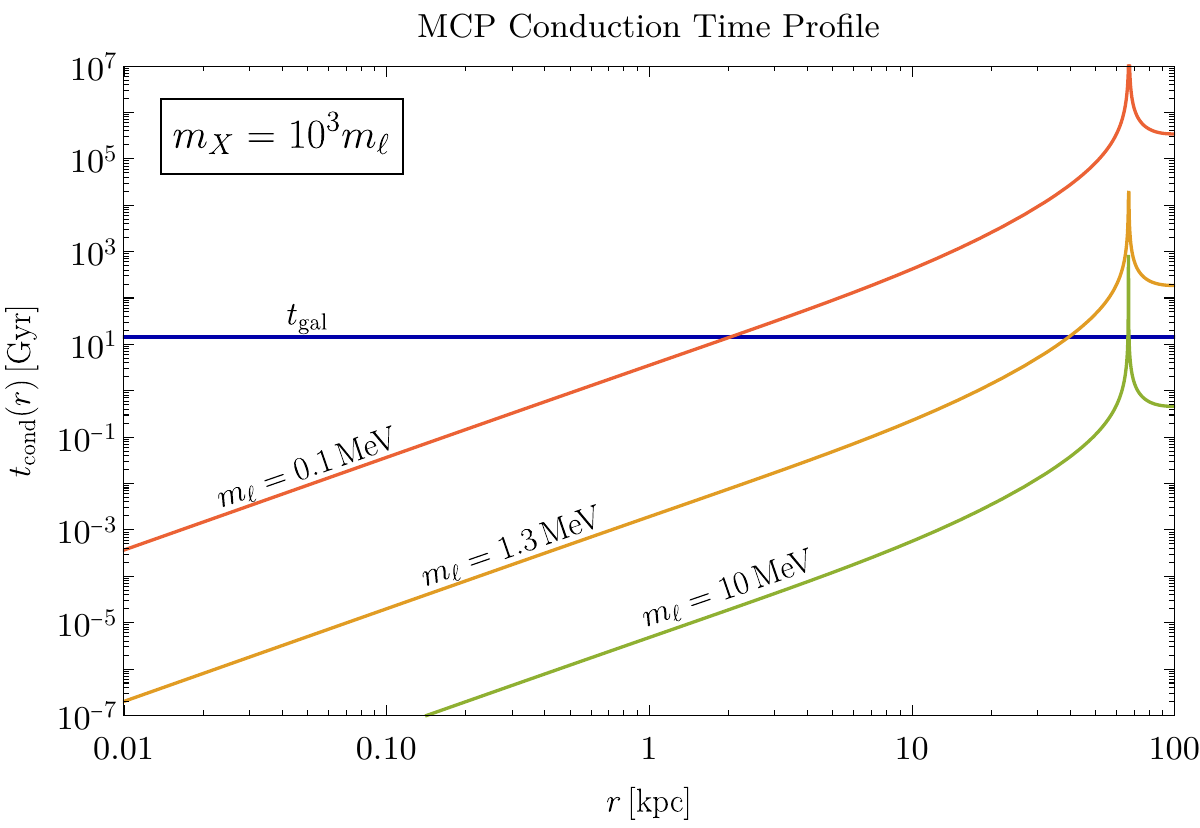}
			\par\end{centering}
		\caption{The conduction timescale in the MCP plasma as a function of the radius, with $t_\textrm{gal}$ for comparison. $t_\textrm{cond}$ was computed in Sec.~\ref{Sec. Thermal Conduction} using the adiabatic halo density and temperature in Eqs.~\eqref{Eq. Explicit Adiabatic Mass-density Profile} and~\eqref{Eq. Explicit Adiabatic Temperature Profile}.\label{Fig. Conduction Timescale Profiles}}
	\end{figure}

    Next, we plot $t_\textrm{cond}$ vs. $t_\textrm{gal}$ as a function of $r$ in Fig.~\ref{Fig. Conduction Timescale Profiles}. The radius at which $t_\textrm{cond}$ diverges is where $\pdv*{T}{t}$ changes sign in Eq.~\eqref{Eq. Conduction Timescale}, which is an artefact of the arbitrary choice of defining the conduction timescale using the initial profiles, while in the dynamical process, the conduction timescale, as defined in  Eq.~\eqref{Eq. Conduction Timescale} will change with time. We compare the two timescales at $r=r_c$, which is chosen to be $r_c=\unit[40]{kpc}$ (see the explanation in Sec.~\ref{Sec. Thermal Conduction} below Fig.~\ref{Fig. Mixed MCP Halo Mass-density and Temperature}), and the profiles for the threshold mass value are again shown in the orange solid lines. 
	
   To complete our discussion, we also consider the mfp of the $\ell$ particles, relevant for the calculation of  the thermalization rate in Sec.~\ref{Sec. l-X Thermalization}. We compute it using the self-collision time $t_{\ell\ell}$~\cite{spitzerplasmabook}
	\begin{equation}\label{Eq. l-l mfp}
		l_\textrm{mfp}^{\ell\ell}=v_\ell t_{\ell\ell}
		=\frac{1}{0.714}\frac{9m_\ell^2\theta_\ell^2}{8\pi\alpha_D^2n_\ell\ln\Lambda},
	\end{equation}
	where $v_\ell=\sqrt{3\theta_\ell}$ is the average speed of $\ell$s. $l_\textrm{mfp}^{\ell\ell}$ is plotted in Fig.~\ref{Fig. Thermalization and l-l mfp Profiles} on the right.
		
	\section{Gravothermal Collapse}\label{App. Gravothermal Collapse}
	
	Self-interacting dark matter (SIDM) halos can in principle form structure, such as in core formation (heat flow from the outer parts of the halo to the center) and core collapse (heat flow from the center outwards). The latter is known as gravothermal collapse, and one might worry if in the parameter space of interest of our model, the MCP halo we've assumed to exist had already collapsed on galactic timescales. In this section we explain why this is not the case. The key property determining core formation/collapse is the rate of heat transfer in the galaxy. It is characterized by the Knudsen number~\cite{Agrawal:2016quu,Ahn:2004xt}
	\begin{equation}
		\textrm{Kn}\equiv\frac{l_\textrm{mfp}}{R_\textrm{halo}},
	\end{equation}
	where $l_\textrm{mfp}$ it the mfp of the DM. $\textrm{Kn}\gg1$ corresponds to almost collisionless dark matter, for $\textrm{Kn}\gtrsim1$ heat transfer is effective and may lead to core formation or collapse, and for $\textrm{Kn}\ll1$ heat conduction is suppressed, inhibiting core formation/collapse. The lighter $\ell$s are more mobile and transfer heat, so the relevant DM mfp is $l_\textrm{mfp}=l_\textrm{mfp}^{\ell\ell}$ in Eq.~\eqref{Eq. l-l mfp}.
	
	For most $m_\ell$ masses considered $\textrm{Kn}\ll1$, so no core formation or collapse is expected. We find $\textrm{Kn}\sim1$ only for masses close to the thermalization limit where the thermalization timescale $t_{\textrm{th}}\lesssim t_\textrm{gal}$. However, for DM with elastic Coulomb-like interactions the timescale of collapse is $\sim330t_{\textrm{th}}$~\cite{1980ApJ...242..765C}, which is longer than the age of the galaxy for these masses. We conclude that in the parameter space of interest gravothermal collapse is either inhibited or hasn't occurred yet.
	
	Our discussion neglects the effects of dissipation, which is justified when the energy loss due to bremsstrahlung is small as explained in Sec.~\ref{Sec. Cooling}. When dissipation is taken into account it can accelerate gravothermal collapse~\cite{Essig:2018pzq}. A detailed analysis of halo evolution with dissipative processes is beyond the scope of this work.

	\end{appendices}
	\bibliographystyle{JHEP}
	\bibliography{boosting_asymmetric_dark_matter_via_thermalization.bib}
\end{document}